\documentclass{aa}     
\usepackage{amssymb}
\usepackage{amsmath}
\usepackage[mathcal]{eucal}
\usepackage{txfonts}
\usepackage{graphicx}

\usepackage[T1]{fontenc}

\usepackage{natbib}

\newcommand{\eqn} [1] {
\begin{equation}
#1
\end{equation}}

\newcommand{\be}{\begin{equation}}
\newcommand{\bea}{\begin{eqnarray}}
\newcommand{\ee}{\end{equation}}
\newcommand{\bc}{\begin{center}}
\newcommand{\ec}{\end{center}}
\newcommand{\bt}{\begin{table}}
\newcommand{\et}{\end{table}}
\newcommand{\eea}{\end{eqnarray}}
\newcommand{\bfig}{\begin{figure}}
\newcommand{\efig}{\end{figure}}
\newcommand{\bfige}{\begin{figure*}}
\newcommand{\efige}{\end{figure*}}

\def\dst {\displaystyle}

\def\gd {{$\gamma$~Doradus}}
\def\gds {{$\gamma$~Doradus~stars}}

\def\teff {T_{\mathrm{eff}}}
\def\logg {{\log\,g}}
\def\rom {{\bar \rho}}

\def\vsini {{v\!\sin\!i}}

\def\rrsol {R/\mathrm{R}_{\sun}}

\def\muHz {{\mu\mbox{Hz}}}
\def\kms {{\mathrm{km}\,\mathrm{s}^{-1}}}
\def\msol {{\mathrm{M}_\odot}}

\def\rsol {\mathrm{R}_{\sun}}

\def\hd {{HD\,48501}}
\def\vaiss {Brunt--Väisälä}

\def\iobs {${\cal I}_{\mathrm{obs}}$}
\def\iteo {${\cal I}_{\mathrm{th}}$}
\def\iobsrot {${\cal I}_{\mathrm{obs,\Omega}}$}
\def\iteorot {${\cal I}_{\mathrm{th},\Omega}$}

\def\ebetao   {\epsilon_{\beta,0}}
\def\ebetarot {\epsilon_{\beta,\Omega}}

\def\dr {\mathrm{d}r}

\def\paperI {paper~I}

\begin{document}

    \title{The Frequency Ratio Method for seismic modelling of \gds}
    \subtitle{II. The role of rotation} 

   \titlerunning{The role of rotation on the FRM}
   \authorrunning{Su\'arez et al.}

   \author{J.C. Su\'arez\inst{1,2*} \and  A. Moya\inst{2} \and S. Mart\'{\i}n-Ru\'{\i}z\inst{1}
           \and P. J. Amado\inst{3} \and A. Grigahc\`ene\inst{1} \and R. Garrido\inst{1}} 	    
    \offprints{J.C. Su\'arez, \email{jcsuarez@iaa.es}\\
              $^*$ Associated researcher at (2)}
    
    \institute{$^1$ Instituto de Astrof\'{\i}sica de Andaluc\'{\i}a (CSIC), CP3004, Granada, Spain\\ 
 	       $^2$ LESIA, Observatoire de Paris-Meudon, UMR8109, France\\
	       $^3$ ESO Office Santiago, Alonso de Cordova 3107, Santiago 19, Chile}

    \date{Received ... / Accepted ...}

\abstract{The effect of rotation on the Frequency Ratio Method 
          \citep{Moya05frmI} is examined. Its applicability
	  to observed frequencies of rotating \gds\ is discussed 
	  taking into account the following aspects: the
	  use of a perturbative approach to compute adiabatic oscillation
	  frequencies; the effect of rotation on the observational
	  \vaiss\ integral determination and finally, the problem
	  of disentangling multiplet-like structures from frequency
	  patterns due to the period spacing expected for high-order gravity
	  modes in asymptotic regime.	  
	  This analysis reveals that the FRM produces
	  reliable results for objects with rotational velocities up 
	  to $70\,\kms$, for which the FRM intrinsic error 
	  increases one order
	  of magnitude with respect to the typical FRM errors
	  given in \citet{Moya05frmI}. 	  
	  Our computations suggest that, given the spherical degree $\ell$ 
	  identification, the FRM may be discriminating for $m=0$
	  modes, in the sense that the method avoids any misinterpretation
	  induced by the presence of rotationally split multiplet-like
	  structures, which reinforces the robustness of the method.
	  However, if $\ell$ is unknown, such
	  discrimination is not ensured. 	  
	  In order to check the FRM in presence of slow-moderate rotation,
	  we have applied it to the three observed frequencies of the
	  slowly rotating ($\vsini=29\,\kms$) \gd\ star 
	  \object{HD\,48501}. 
}
\maketitle


\section{Introduction}

In \citet{Moya05frmI} (hereafter \paperI), we presented a method for obtaining 
asteroseismic information of \gds\ when at least three oscillation frequencies 
are observed. This method, from now on called FRM (Frequency Ratio Method),
is particularly useful for improving our knowledge on the pulsational behaviour
of \gds\ through the following aspects: first, the identification of the radial
order $n$ and the mode degree $\ell$ of observed frequencies and second, the possibility
of constraining models from an additional stellar quantity, ${\cal I}$ (the 
integral of the buoyancy or \vaiss\ frequency weighted over the stellar
radius along the radiative zone). 
Such constraints can be interpreted as an increase of the classical
observables, 
\eqn{
{\cal P}={\cal P}(Z, g, L, \teff; \frac{f_{{\rm o},i}}{f_{{\rm o},j}}, {\cal I}),\nonumber
}
where $Z$ represents the relative metal abundance; $g$ the surface gravity,
$L$ the luminosity; $\teff$ the effective temperature, and finally, $f_{{\rm o},i}$ 
and $f_{{\rm o},j}$
the observed frequencies. In \paperI, the method was applied to the \gd\ star 
\object{HD\,12901}, for which
a significant reduction of possible representative models of the star was 
achieved.

The FRM, based on the first-order
asymptotic $g$-mode expression given by \citet{Tassoul80} (see Eq.~1 in \paperI),
is built under the assumptions of adiabaticity and non rotation.
However, recent spectroscopic measurements of 59 \gd\ candidates
\citep{Mathias04} show rotational velocities ($\vsini$ measurements) varying
from 10 to $160\,\kms$. This corresponds to rotational 
periods around 0.4 to 4--5 days, in the limit of applicability of a
perturbative analysis used here to compute adiabatic oscillation frequencies. 

It has been shown by \citet{MiHer99} and references
therein, that photometric parameters are sensitive to rotation.
In particular, it modifies the determination of the position of 
the stars in the HR diagram. Since the FRM results are sensitive
to such variations, the rotation must thus be included in the
set of \emph{working} parameters 
\eqn{
{\cal P}={\cal P}_\Omega(Z, g, L, \teff; \frac{f_{o,i}}{f_{o,j}}, {\cal I}, \Omega),\nonumber
}
where $\Omega$ represents the rotational velocity of the star. 

Concerning the theoretical adiabatic oscillations, special care must be
taken when the oscillation frequencies are of the order of magnitude of the rotational
frequency of the star, i.e. when $\sigma_{\Omega}\sim\sigma/3$. For such cases the
perturbative method for computing adiabatic frequencies may start to fail, and 
other non-perturbative theories should be used. In this context, 
several works can 
be found in the literature: \citet{Dintrans00}, \citet{Rieutord02}, 
\citet{Dintrans04}, in which non-radial gravity modes of a typical
\gd\ star are studied by means of polytropic models. 

In addition, the effects of rotation on the oscillation spectra 
may introduce additional uncertainties to the FRM. 
The problem arises when the
classical $g$-mode frequency pattern may be misidentified with multiplet-like 
splitting rotationally induced structures.
Ideally, when considering complete asymptotic spectra of \gds,
in a first approximation, the asymptotic $g$-mode pattern, whose
periods are regularly spaced, can be easily disentangled from 
multiplet-like structures, which are regularly spaced in frequency.
However in practice, unfortunately only a few oscillation modes
are observed, and therefore, such direct discrimination
is not accomplishable.

The paper is organised as follows: In Sect.~\ref{sec:models}, the
equilibrium models are described. The domain of validity of the FRM
when applied to rotating \gds\ is analysed in Sect.~\ref{sec:errors}.
In Sect.~\ref{sec:triplets-game}, the  problem of multiplet-like structures
in the framework of the FRM is discussed. An application of 
the FRM for the rotating \gd\ star HD\,48501 is given in 
Sect.~\ref{sec:applic}. Finally, conclusions are summarised in
Sect.~\ref{sec:conclu}. 

\section{The models  \label{sec:models}}

For the purpose of constructing representative models of \gds\ the evolutionary
code CESAM \citep{Morel97} is particularly adapted. Models are
built with a precision optimised to compute
oscillations, i.e. around 2000 mesh points for the equilibrium model mesh grid,
given in the basis of B-splines.
To take into account
a first order effect of the rotation, equilibrium equations are modified
in the manner described in \citet{KipWeig90}. In particular, the 
spherical symmetric contribution of the centrifugal acceleration
is included by means of an effective gravity
\eqn{g_{\mathrm{eff}}=g-{\cal A}_{c}(r)\,,\nonumber}
where $g$ represents the local gravity component, $r$ the radial distance 
and, 
\eqn{{\cal A}_{c}(r)=\frac{2}{3}\,r^2\,\Omega\nonumber}
the centrifugal acceleration of matter elements at a distance $r$ 
from the centre of the star. This spherically symmetric 
contribution of the rotation does not change the 
shape of the hydrostatic equilibrium equation. Although, the non-spheric
components of the centrifugal acceleration are not considered, they are
included as a perturbation in the oscillation computation. 
The total angular momentum of models is assumed to
be globally conserved along the evolution of the star.

The physical parametrisation has been adapted for intermediate mass
stars (see \paperI). In particular, all equilibrium models
used in the present work are computed with : 
$\alpha_{ML}=l_{m}/\mathrm{H}p=1.8$, for the convective efficiency
and $d_{ov}=l_{ov}/\mathrm{H}p=0.2$ 
for the mixed core overshooting. The $\mathrm{H}p$ corresponds to the local 
pressure scale-height; 
$l_{m}$ and $d_{ov}$ represent the mixing length and the inertial penetration 
distance of convective bulbs respectively.

\section{The error determination  \label{sec:errors}}

The study of the influence of rotation on the FRM constitutes a 
rather complex task. 
We aim at identifying possible uncertainties 
coming from rotation effects and their relevance for the FRM.
Such analysis is here focused in three important aspects: first, the regime of application of 
the perturbative method
for obtaining theoretical adiabatic frequencies. Second,
the quantification of the error committed when considering
the classical analytical expression for asymptotic $g$ modes
to study frequency ratios of rotating \gds.
And third, the influence of rotation on the \vaiss\ integrals ${\cal I}$.
\begin{figure*}
 \begin{center}
   \includegraphics[width=9cm]{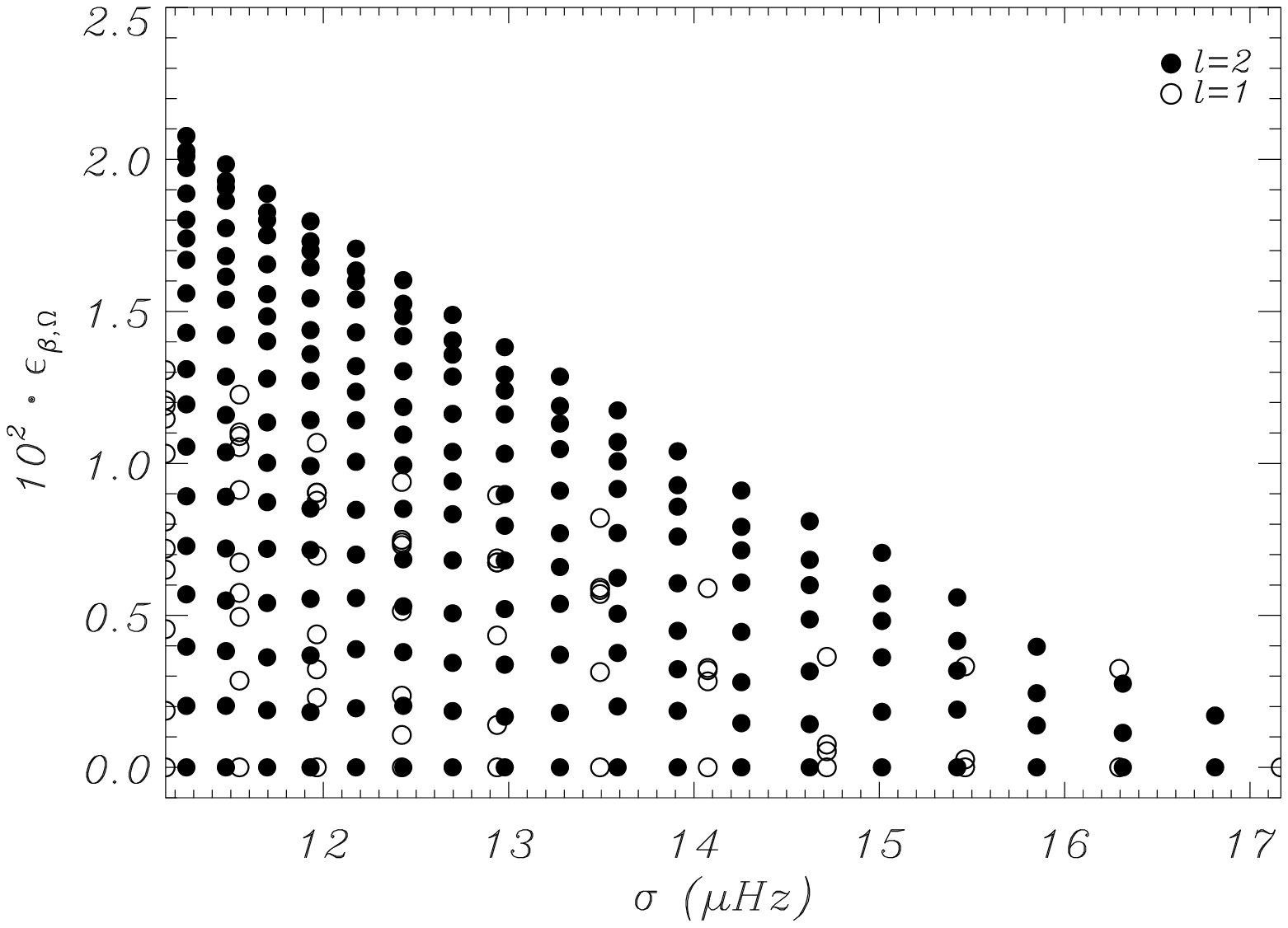}
   \includegraphics[width=9cm]{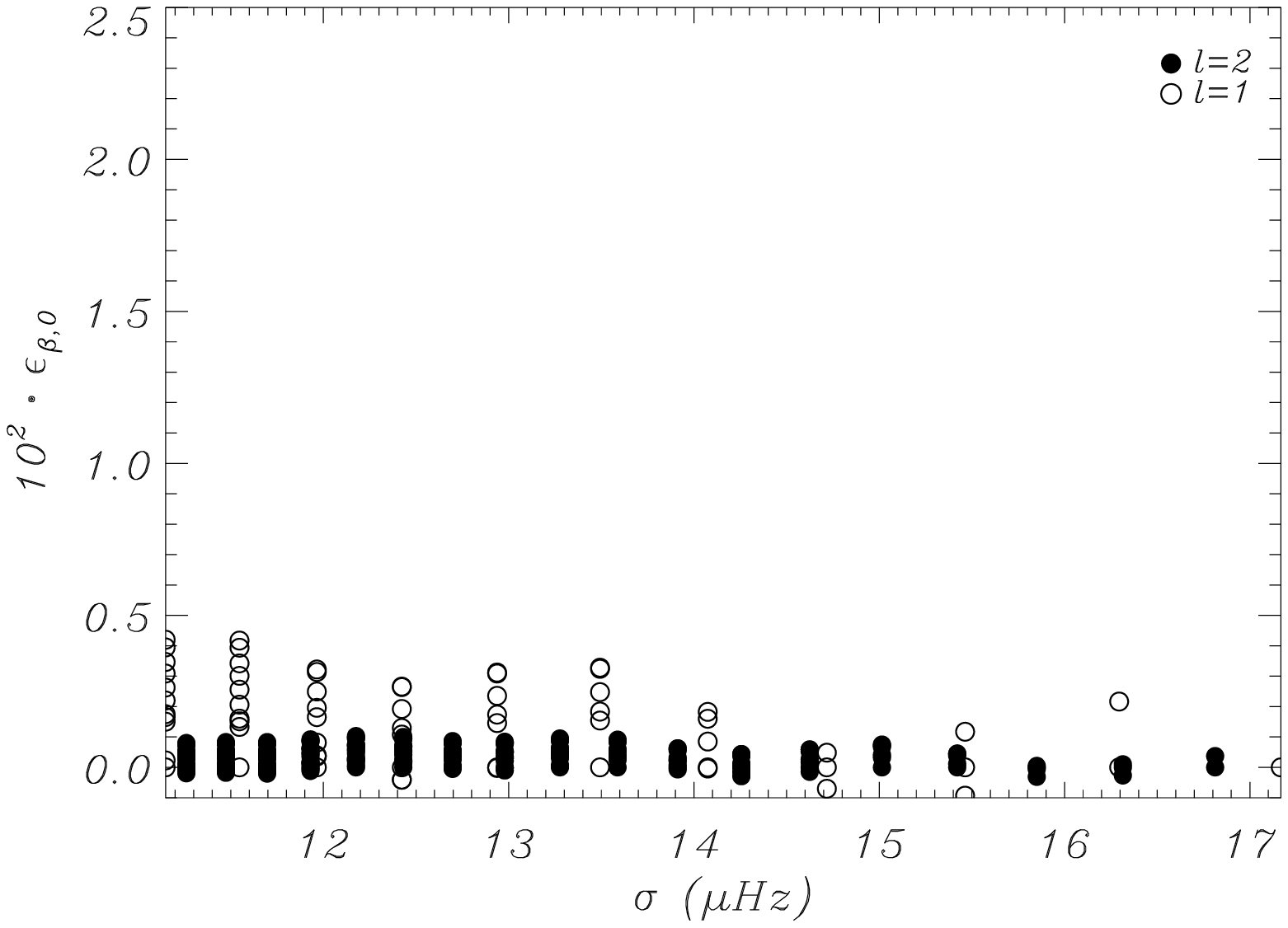}
   \caption{Relative errors of the asymptotic frequencies as compared to the
            theoretical predictions for a complete oscillation spectrum
	    obtained from a model with $\Omega\sim50\,\kms$. 	    
	    Left panel
            represent relative errors obtained from oscillations frequency ratios 
            corrected for the effect of rotation. Right panel shows errors
	    without correcting for the effect of rotation.}
   \label{fig:epsbetarot_err}
 \end{center}
\end{figure*}

\subsection{The perturbative approach for adiabatic oscillation frequencies 
\label{ssec:pert-approach}}

The limit for using the perturbative approach to compute adiabatic oscillation frequencies 
is given by the relation $\sigma/\sigma_\Omega\gg1$, i.e. when the rotational 
frequency of the star ($\sigma_\Omega$) is negligible with respect to the
the oscillation frequency of modes ($\sigma$). 
In the asymptotic regime, $g$ modes are 
characterised 
by their low frequency (high radial order $n$).
According to \citet{Dintrans00}, the most perturbed modes by rotation are 
likely those with $\sigma\lesssim 2\,\sigma_\Omega$, where $2\,\sigma_\Omega$ corresponds to the
Coriolis frequency. Considering the typical rotational periods of \gds, low frequency
gravity modes in asymptotic regime are strongly affected by rotation.

The present analysis is thus restricted to slow and moderately rotating
\gds, that is,
those with rotational velocities smaller than 50-$70\,\kms$. Considering representative
models with masses in the range of $1.4\,$--$\,1.7\msol$, the limits of the perturbative
approach can be established to have $n$=[30, 40] for modes with $\ell\leq2$,
expected to have significant amplitude not smeared out by cancellation effects.

\subsection{The analytic expression for the asymptotic $g$ mode frequencies 
\label{ssec:analexpr}}

It is convenient to remind that the analytic expression for 
the asymptotic $g$ modes
used in the FRM is obtained under the hypothesis of non-rotation (see \paperI).
We attempt here to determine the accuracy of this expression when 
rotation effects are considered.
In this framework, it is conceivable to define an equivalent form of the analytic
expression for the asymptotic $g$ modes as:
\eqn{\sigma_{n,\ell,\Omega}^{\,a}\equiv\,f(n,\ell)\,{\cal I}_\Omega\,\,
    \label{eq:defsigmarot}}
where the analytic form of ${\cal I}_\Omega$ integrals remains
identical to  ${\cal I}$ (see \paperI), and whose values are assumed to be different to those
obtained without considering rotation effects. The $f(n,\ell)$ functions
are a priori unknown and assumed to be only dependent on $\ell$
and $n$ numbers. 
Similarly as done in \paperI, two given $g$ modes, 
$\sigma_{\alpha_1,\Omega}$ and $\sigma_{\alpha_2,\Omega}$, are considered, where  
$\alpha_1$ and $\alpha_2$ represent the pairs $({n_1},{\ell_1})$ and
$({n_2},{\ell_2})$ respectively. Assuming the same
mode degree $\ell_1=\ell_2=\ell$, we make the hypothesis 
that their eigenfrequencies are
approximated by Eq.~\ref{eq:defsigmarot}. If 
the star rotates slowly (cf. Sect.~\ref{ssec:pert-approach}), 
$f(n,\ell)$ can be assumed to be: 
\eqn{f(n,\ell)\sim\dst\left(\frac{\dst\sqrt{\ell(\ell+1)}}{(n+1/2)\,\pi}\right)_\Omega\,,
\label{eq:approx_fn}}
and Eq.~\ref{eq:defsigmarot} can thus be rewritten as follows:
\eqn{\sigma_{n,\ell,\Omega}^{\,a}=
     \left(\frac{\dst\sqrt{\ell(\ell+1)}}{(n+1/2)\,\pi}\right)_\Omega\,
     {\cal I}_\Omega\,.\label{eq:approx_sigmarot}}
In this way, the analytic form proposed by \citet{Tassoul80} for non-rotating
stars is kept and consequently, the formal structure of the FRM
remains unaltered.
This allows us to relate $\sigma_{\alpha_1,\Omega}$ and $\sigma_{\alpha_2,\Omega}$ through the
following ratio
\eqn{
\dst\left(\frac{\sigma_{\alpha_1}}{\sigma_{\alpha_2}}\right)_\Omega\approx
\dst\left(\frac{n_2+1/2}{n_1+1/2}\right)_\Omega\,.\label{eq:sigma12ratio_rot}}
To obtain this expression, we proceed similarly as in \paperI\ (Sect.~3). 
It can be shown that the small differences in the outer turning point location 
can be considered as negligible as far as the calculation of the 
\vaiss\ frequency integral is concerned.

Provided theoretical frequencies and the corresponding
$n's$ ratio estimate for a given model, the accuracy of 
Eq~\ref{eq:sigma12ratio_rot} can be defined as 
\eqn{\ebetarot=\dst\left(\frac{\sigma_i}{\sigma_j}\right)_\Omega-
          \dst\left(\frac{f(n_j,\ell)}{f(n_i,\ell)}\right)_\Omega
\label{eq:def_epsbetarot}}
Figure~\ref{fig:epsbetarot_err} (left panel) shows such errors obtained from a 
theoretical oscillation spectra of a $1.40\,\msol$ model. 
It can be noticed that errors reach 0.013 
for $\ell=1$ modes, and 0.021 for
$\ell=2$ modes, for rotational velocities around 50--$60\,\kms$. 
These results represent around one order of magnitude
larger than the $\ebetao$ errors (right panel), obtained without taking into account
corrections for rotation in frequencies. 
As a consequence, the 
quantity of valid sets of natural numbers ($n_i$) increases 
(see \paperI, Sect.~3).
Such increase can be of the order of 10 additional sets for 
$\ell=1$ modes, and even 100 sets for $\ell=2$ modes.
Nevertheless, as explained in \paperI\, each additional 
possible set, implies an extra constraint for 
the \vaiss\ integrals which must be fulfilled.

\subsection{The effects of rotation on the ${\cal I}$ integrals \label{ssec:iteo-rot}}

This study must be completed with the analysis of the behaviour
of ${\cal I}$ integrals when rotation is taken into account. 

As explained in \paperI, in the FRM framework the \vaiss\ frequency integrals are
calculated in two different manners: from equilibrium models and directly
from frequency ratios as deduced from the analytic expression for asymptotic
$g$ modes (Eq.~\ref{eq:approx_sigmarot}).
\begin{figure}
 \begin{center}
   \includegraphics[width=9cm]{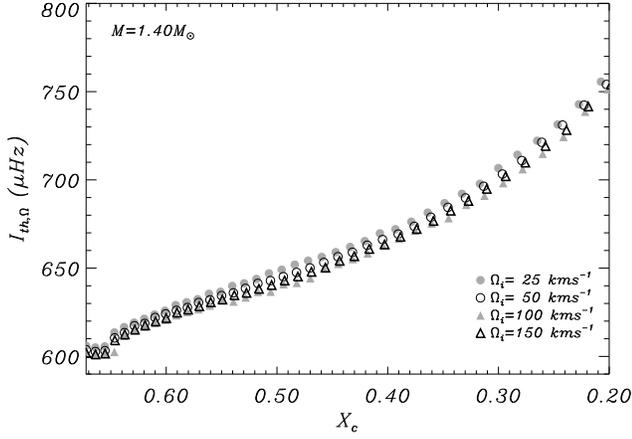}
   \caption{Variation of theoretical \vaiss\ frequency integral  
            as a function of the evolutionary stage (represented
            by the central Hydrogen fraction $X_c$) for a $1.4\,\msol$ model
	    computed with solar metallicity. Symbols represent
	    models with different initial (ZAMS) rotational 
	    velocity.}
   \label{fig:iteorot_xc}
 \end{center}
\end{figure}
\begin{figure}
 \begin{center}
   \includegraphics[width=9cm]{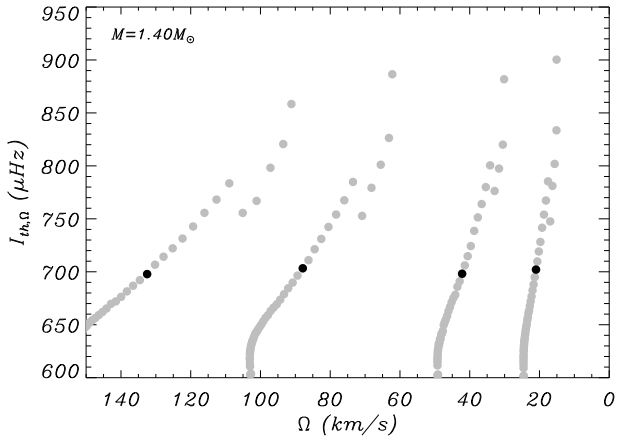}
   \caption{Theoretical \vaiss\ integral as a function of the rotational velocity
            $\Omega$.
            Filled grey circles represent the same $1.4\,\msol$ models evolved with
	    the same initial rotational velocities considered in 
	    Fig.~\ref{fig:iteorot_xc}.
	    Filled black circles represent selected models with the \iteo\ values
	    closest to $700\,\muHz$ (see details in the text).}
   \label{fig:I-omega}
 \end{center}
\end{figure}
The characteristics of pseudo-rotating models 
(cf. Sect.~\ref{sec:models}) allow us to consider equivalent internal
structure for rotating \gds\, only modified by an effective gravity.
Therefore, the \vaiss\ frequency integrals can be calculated as
for non-rotating models, that is 
\eqn{{\cal I}_{{\rm th}}=\int_{r_a}^{r_b} \frac{N}{r}\dr
      ={\cal I}_{{\rm th},\Omega}
      \equiv\frac{\sigma_{n,\ell,\Omega}^{\,a}}{f(n,\ell)}\,. 
      \label{eq:defIobs}}
These integrals
are dependent mainly on the mass and metallicity of models 
(Fig.~5 in \paperI). For rotating \gds\ it is plausible to 
consider also a $\Omega$ dependence
\eqn{{\cal I}_{\mathrm{th},\Omega}={\cal I}_{\mathrm{th}}(M,Z,\Omega)\,.
     \label{eq:def_iteorot}}
For the purpose of investigating the influence of rotation on \vaiss\ integrals, we
consider $1.4\,\msol$ pseudo-rotating models as described in
Sect.~\ref{sec:models}. 
\begin{table}
   \caption{Characteristics of selected models in Fig.~\ref{fig:I-omega},
            presenting a common \iteo$\,\sim\!700\,\muHz$, and evolved from
	    the initial (at ZAMS) rotational velocities $25, 50, 100$ and $150\,\kms$ 
	    ($\Omega_{1}$, $\Omega_{2}$, $\Omega_{3}$
	    and $\Omega_{4}$ respectively). The following
	    stellar parameters are given: $\teff$, the effective temperature (in 
	    logarithmic scale); $g$, the surface gravity (in a logarithmic
	    scale); $\rom$, the mean stellar density (in g\,cm$^3$); $\rrsol$, the stellar
	    radius (in solar units); $X_c$, the central Hydrogen fraction;
	    $\Omega$, the rotational velocity (in $\kms$), and finally, \iteo,
	    the theoretical \vaiss\ frequency integral (in $\muHz$).}
   \begin{center}
      \begin{tabular}{rrrrr} \hline\hline
      \noalign{\medskip}
          & $\Omega_{4}$ & $\Omega_{3}$ & $\Omega_{2}$ & $\Omega_{1}$ \\ 
   \noalign{\medskip}
   \hline  
   \noalign{\medskip}	
  $\log\teff$ &   3.81   &   3.81   &   3.82  &   3.82 \\
      $\logg$ &   3.96   &   3.96   &   3.97  &   3.97 \\ 
      $\rom$  &   3.79   &   3.75   &   3.88  &   3.84 \\     
    $\rrsol$  &   2.18   &   2.19   &   2.17  &   2.17 \\           
       $X_c$  &   0.32   &   0.30   &   0.29  &   0.28 \\
     $\Omega$ & 132.50   &  87.87   &  42.18  &  20.99 \\
      \iteo   & 697.81   & 703.31   & 698.15  & 702.08 \\         
    \noalign{\smallskip}
      \hline 
      \noalign{\smallskip}
      \end{tabular}
      \label{tab:isoImodels}
   \end{center}
\end{table}
Models are evolved from the ZAMS
to the subgiant region with four different 
initial rotational velocities: 25, 50, 100 and $150\,\kms$. 
For these models, the evolution of \iteorot\ 
is depicted in Fig.~\ref{fig:iteorot_xc} as a function
of the core hydrogen fraction $X_c$. Small variations
($\lesssim 10\,\muHz$) are observed. As expected, the effect of 
considering an effective gravity does not modify the 
behaviour of the theoretical \vaiss\ integrals along
the stellar evolution.

In practice, initial rotational velocities are unknown. Therefore,
it is worth to analyse the characteristics of models 
presenting a similar \vaiss\ integral but having been evolved
with different initial rotational velocities. 
In Fig.~\ref{fig:I-omega}, the evolution of \iteorot\ integrals is shown
as a function of the rotational velocity. Each track corresponds
to evolved models (from left to right) with $\Omega_{\mathrm{ZAMS}}=$150, 100, 50
and $25\,\kms$. From each set of models, those with 
\iteorot$\sim700\,\muHz$ (black circles) have been arbitrarily selected.
The analysis of the corresponding stellar parameters
given in Table~\ref{tab:isoImodels} reveals similar characteristics
for the four models. Small differences are mainly due to the
different evolutionary stage of models. 
\begin{figure*}
 \begin{center}
   \includegraphics[width=9cm]{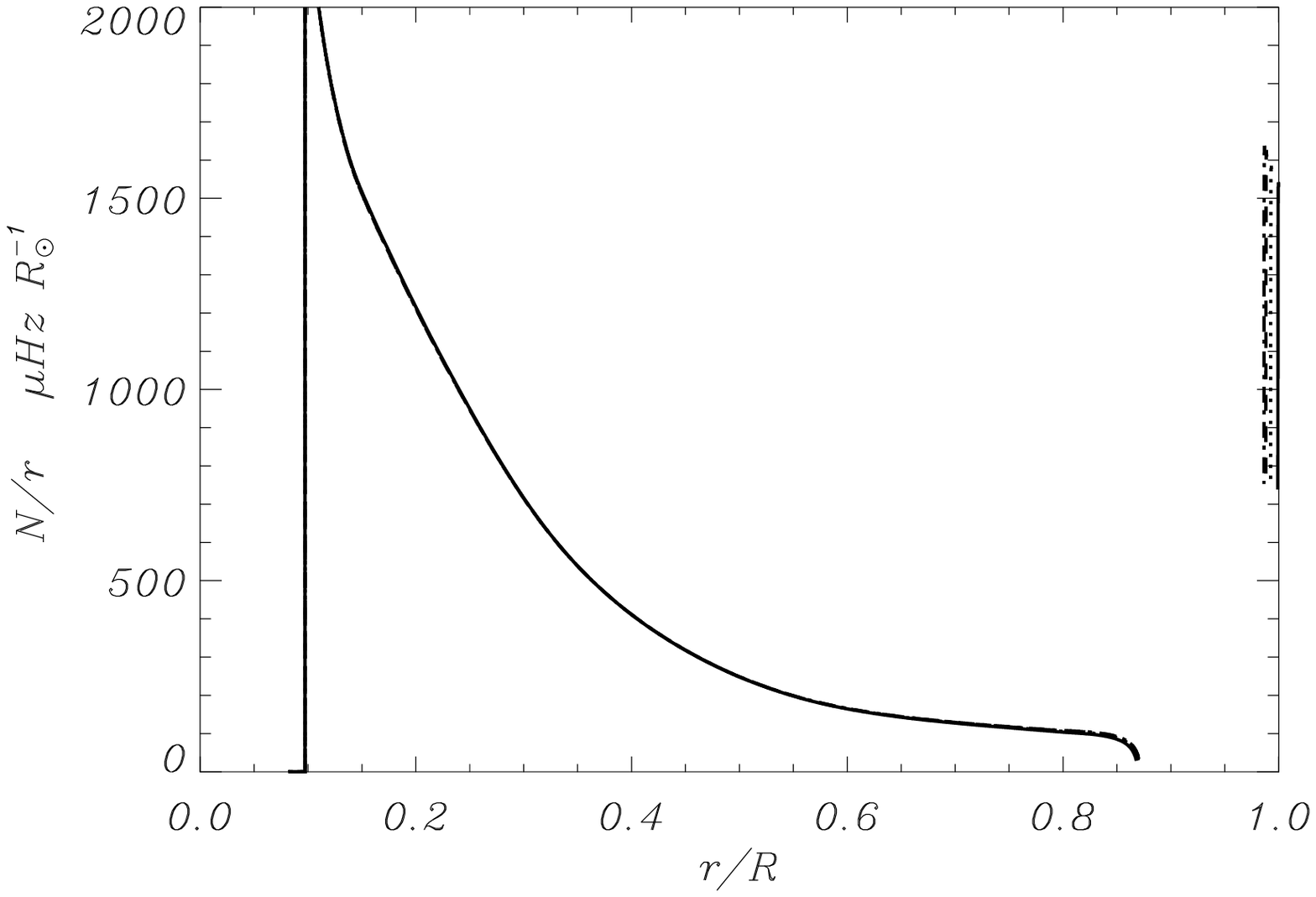} \hspace{-1cm}
   \includegraphics[width=9cm]{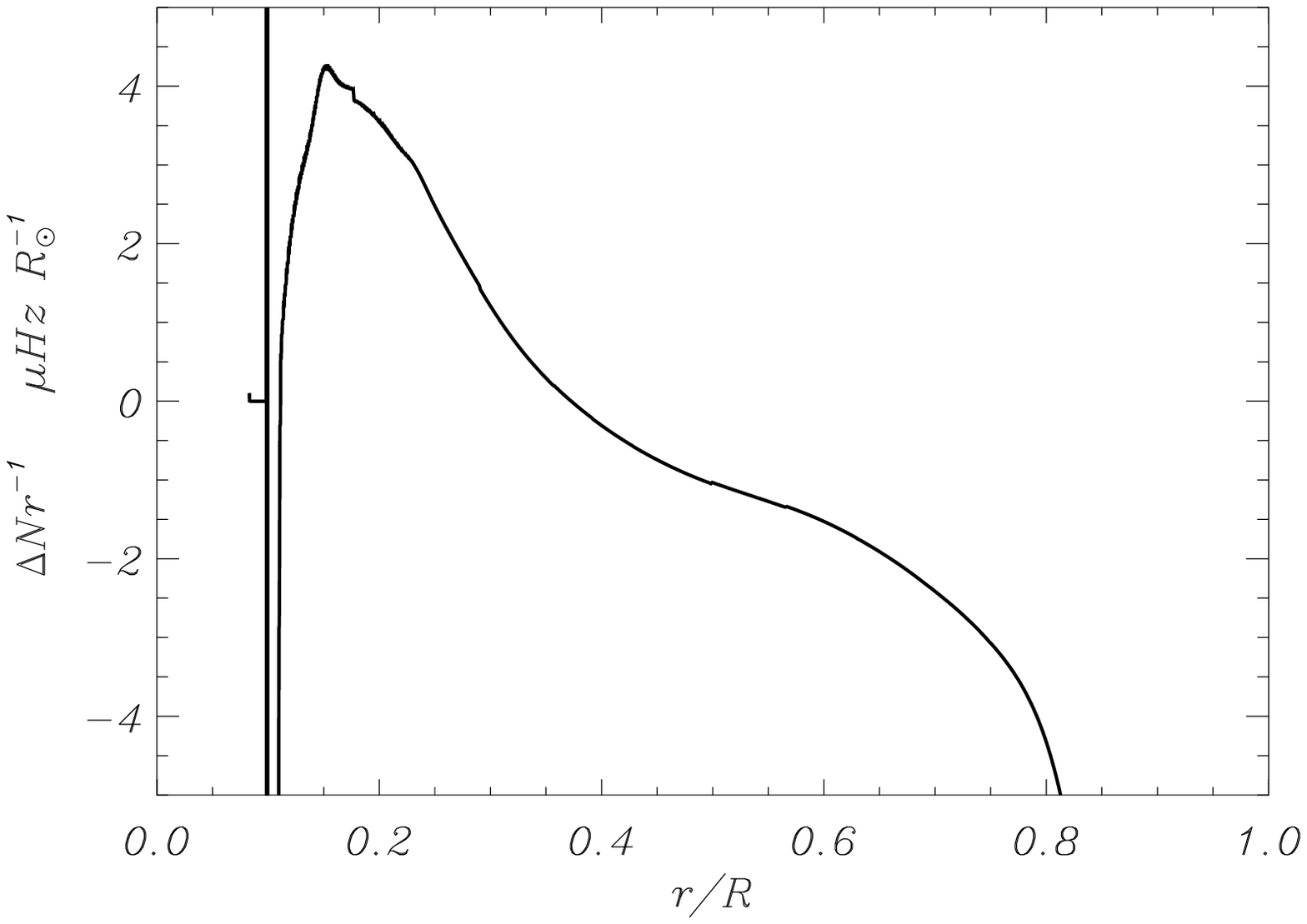}
   \caption{\vaiss\ frequency weighted over the stellar radius (in solar units) as a function 
            of the radial distance $r$ (normalised to the stellar radius R), for
	    the model with the rotational velocity $\Omega_1$ (see Table~\ref{tab:isoImodels}. 
	    Right panel shows
	    the relative difference between the \vaiss\ frequency 
            weighted over the stellar radius of models 
            with $\Omega_1$ and $\Omega_4$ respectively (see Table~\ref{tab:isoImodels}). }
   \label{fig:nr-r_diffN-r}
 \end{center}
\end{figure*}
In particular, 
the $\pm5\,\muHz$ dispersion in \iteorot\ values are due to
slight differences in their integrands (Fig.~\ref{fig:nr-r_diffN-r}). 
As expected, rotation does
not discriminate models with respect to their \iteorot. Nevertheless
this ensures the coherence of the FRM, since rotation does not
introduce additional uncertainties from the point of view of
the \iteo\ behaviour.

Le us consider now the observed \vaiss\ frequency integrals.
The analysis of uncertainties of such integrals is not
trivial. According to Eq.~\ref{eq:defsigmarot}, these integrals
can be expressed as
\eqn{{\cal I}_{{\rm obs},\Omega}=\frac{\sigma_{n,\ell,\Omega}}{f(n,\ell)}\, .
     \label{eq:isol_iobs}}
From this expression and Eq.~\ref{eq:defIobs}, it is possible 
to define the error associated to $\epsilon_{\sigma,\Omega}$ as
\eqn{\epsilon_{\sigma,\Omega}\equiv
     \dst\frac{|\sigma_{n,\ell,\Omega}-
     f(n,\ell)\,{\cal I}_{{\rm th},\Omega}|}{\sigma_{n,\ell,\Omega}}\,.
     \label{eq:deferrorSigma}}
Respectively, the error associated to the \vaiss\ frequency integral
can be written as     
\eqn{\epsilon_{{\cal I},\Omega}\equiv
     \dst\frac{|{\cal I}_{{\rm th},\Omega}-
               {\cal I}_{{\rm obs},\Omega}|}{{\cal I}_{{\rm th},\Omega}}\,.
     \label{eq:deferrorI}}
Multiplying both sides of 
Eq.~\ref{eq:deferrorI} by $f(n,\ell)$\footnote{This operation is licit since
$f(n,\ell)>0$ by construction.}, and then considering the Eqs.~\ref{eq:isol_iobs} 
and Eq.~\ref{eq:deferrorSigma}, both errors can be related as follows:
\eqn{\dst\frac{\epsilon_{\sigma,\Omega}}{\epsilon_{{\cal I},\Omega}}=
          \dst\frac{f(n,\ell)\,{\cal I}_{{\rm th},\Omega}}{\sigma_{n,\ell,\Omega}}\,.
	  \label{eq:eps_ratios}}
According to Eq.~\ref{eq:deferrorSigma}, $\epsilon_{\sigma,\Omega}$ can be
discussed as a function of the sign of 
\begin{equation}
  1-\dst\frac{f(n,\ell)}{\sigma_{n,\ell,\Omega}}\,{\cal I}_{{\rm th},\Omega}=
     \begin{cases}
      +\epsilon_{\sigma,\Omega} & 
        \text{if $\dst\frac{f(n,\ell)}{\sigma_{n,\ell,\Omega}}\,{\cal I}_{{\rm th},\Omega}>1$}\\
      -\epsilon_{\sigma,\Omega} & 
        \text{if $\dst\frac{f(n,\ell)}{\sigma_{n,\ell,\Omega}}\,{\cal I}_{{\rm th},\Omega}<1$}  
 \end{cases}
 \label{eq:errorcond}
\end{equation}
where the trivial case of $\epsilon_{\sigma,\Omega}=0$ is obviously omitted.
The right hand side of Eq.~\ref{eq:eps_ratios} can thus be written as
\eqn{\dst\frac{f(n,\ell)\,{\cal I}_{{\rm th},\Omega}}{\sigma_{n,\ell,\Omega}}=
      1\pm\epsilon_{n,\ell,\Omega}\,,\label{eq:fI_sigma}}
This allows us to quantify the error in ${\cal I}_{{\rm obs},\Omega}$ integrals
as a function of $\epsilon_{\sigma,\Omega}$ by replacing the left hand side
of Eq.~\ref{eq:eps_ratios} in Eq.~\ref{eq:fI_sigma}, yielding
\eqn{\epsilon_{{\cal I},\Omega}=
      \dst\frac{\epsilon_{\sigma,\Omega}}{1\pm\epsilon_{\sigma,\Omega}}\,,
     \label{eq:Ierr_isol}}
This means that $\epsilon_{{\cal I},\Omega}$ errors can thus be 
considered of the same order as $\epsilon_{\sigma,\Omega}$. Such result
was somehow anticipated in \paperI\ from numerical computations. Here,
although the relation given by Eq.~\ref{eq:Ierr_isol} has been obtained
for $\sigma_{n,\ell,\Omega}$ and ${\cal I}_{{\rm obs},\Omega}$ errors, it
can be easily shown that is generally valid whatever considered
errors and $f(n,\ell)$ values. Similarly as it happens for non-rotating models,
$\epsilon_{{\cal I},\Omega}\sim\,10^{-3}$ which implies an
error of $10^{-3}$ in $\epsilon_{\sigma,\Omega}$.

\section{The problem of multiplet-like structures \label{sec:triplets-game}}

When rotational splittings are of the same
order as the period spacing of high radial order
$g$ modes, i.e.,
\eqn{|\sigma_{n,\ell,m}-\sigma_{n,\ell,m^\prime}|\sim
     |\sigma_{n,\ell,0}-\sigma_{n',\ell,0}|\,,
     \label{eq:condtrip}}
the FRM predictions may fail. In particular, rotationally split
multiplets (or any of their $m\neq0$ components) may be confused with the large
separation pattern expected for high-order gravity modes
in asymptotic regime.

In order to investigate this, an \emph{observed} oscillation spectrum of a 
given rotating \gd\ star has been simulated. This simulation has been
performed using a pseudo-rotating model (as described in 
Sect.~\ref{sec:models}) with a mass of 
$1.70\,\msol$, and a rotational velocity of $\Omega_0=26.8\,\kms$.
Computed with solar metallicity, it is located in the 
main sequence (age 1.3 Gyr). It presents an effective temperature
and a surface gravity ($\logg$, cgs) of 
$6918\,\mathrm{K}$ and 3.94 respectively. Its theoretical
\vaiss\ frequency is \iteorot$=667.8\,\muHz$.

Adiabatic eigenfrequencies have been computed from this model
with the oscillation code FILOU 
\citep[see][]{filou}. In this code, effects of rotation up to second 
order on the oscillation frequencies have been included \citep{SuaThesis} 
following \citet{Soufi98}. 
Due to the nature of the observed frequencies, 
the theoretical oscillation computations are extended to
high order $g$ modes ($-40\leq\,n\,\leq 1$), limiting our 
investigations to modes with degrees $\ell\leq2$, since 
cancellation effects are important for the visibility of modes
with $\ell\geq3$.
\begin{table}
   \caption{List of possible ($n$, ${\cal I}$) identifications provided by
            the FRM applied to the theoretical frequencies belonging to
	    the rotational ($n=22,\ell=1$) triplet (more details in text). 
	    The first three columns represent
	    the resulting radial order identification. Last two, list the
	    corresponding \emph{observed} ${\cal I}$ when assuming $\ell=1$ and 2
	    respectively. }
   \begin{center}
      \begin{tabular}{rrrrr} \hline\hline
      \noalign{\medskip}
      $n_1$ & $n_2$ & $n_3$ &\iobs ($\ell=1$) & \iobs ($\ell=2$)\\ 
      \noalign{\smallskip}
      \hline 
      \noalign{\smallskip}
        26 & 23 & 21 &  704.379 & 406.673\\
        27 & 24 & 22 &  734.353 & 423.978\\
        28 & 25 & 23 &  764.326 & 441.284\\
        36 & 32 & 29 &  974.141 & 562.421\\
        37 & 33 & 30 & 1004.115 & 579.726\\
        38 & 34 & 31 & 1034.089 & 597.031\\
        39 & 35 & 32 & 1064.062 & 614.336\\
        40 & 36 & 33 & 1094.036 & 631.642\\
        42 & 37 & 34 & 1124.009 & 648.947\\
        43 & 38 & 35 & 1153.983 & 666.252\\
        44 & 39 & 36 & 1183.957 & 683.557\\
        45 & 40 & 37 & 1213.930 & 700.863\\
      \noalign{\smallskip}
      \hline 
      \end{tabular}
      \label{tab:ident_triplet}
   \end{center}
\end{table}

To cover all the possible situations where multiplet-like structures
may \emph{contaminate} the FRM results, the following scenarios
have been explored:

\begin{enumerate}
  \item the observed frequencies belong to any rotational 
        multiplet; 
  \item they belong to modes with different radial order $n$ and 
        azimuthal orders $m$; 
  \item they are identified as modes with different ($n$) but with
        $m=0$.
\end{enumerate}       

\subsection{Scenario 1\label{ssec:scen1}}
For the first scenario, let us examine, for instance, the 
\emph{observed} $(n,~\ell)=(22,~1)$ triplet with frequencies (in $\muHz$) 
\eqn{\sigma_{-1}=14.756,\;\;\sigma_{0}=13.493,\;\;\sigma_{+1}=12.045,
     \nonumber\label{eq:game_triplet}}
and with frequency ratios     
\eqn{\frac{\sigma_{+1}}{\sigma_{0}}=0.8927,\;\;\frac{\sigma_{0}}{\sigma_{-1}}=0.9144,\;\;
      \frac{\sigma_{+1}}{\sigma_{-1}}=0.8163\, .
      \nonumber\label{eq:game_triplet_ratios}}
For these frequency ratios, the FRM predicts a list of possible 
($n$, ${\cal I}$) identifications indicated in Table~\ref{tab:ident_triplet}. 
As discussed in 
Sect.~\ref{ssec:analexpr}, this list has been constructed by assuming
an error of $\pm 1.3\,10^{-2}$ for $\epsilon_{\beta}$. An estimate of
\iobsrot\ integrals
is also given for the two most relevant mode degree values: $\ell=1$ and 2.
The \emph{observed} set is obviously not predicted since $n_1=n_2=n_3$.
When analysing the list of possible \iobsrot\ integrals, 
predictions exceed the error limit (around 2\%) when
assuming $\ell=1$. That is, for this known value of the
mode degree, none of the potential sets would be misinterpreted. 
The same exercise repeated with other 
$\ell=1$ multiplets within the same spectral range yields similar
results.
In contrast, the list of \vaiss\ integrals
predicted assuming $\ell=2$ contains \iobsrot\ values close to
the theoretical one. In particular, the set 
$(n_1, n_2, n_3)=(43, 38, 35)$ is related to  
\iobsrot($\ell=2$)=$666.252\,\muHz$, which represents
an error of 0.23\% approximately with respect to the
\iteo. This case would thus be relevant when no
additional information on the mode degree is provided. In such
a situation, the discrimination between a possible multiplet-like
structure and asymptotic $g$ mode pattern is not ensured.
\begin{table}
   \caption{Same as Table~\ref{tab:ident_triplet} but obtained
            for the case of modes with different radial and
	    azimuthal orders (more details in the text). }
   \begin{center}
      \begin{tabular}{rrrrr} \hline\hline
      \noalign{\medskip}
      $n_1$ & $n_2$ & $n_3$ &\iobs ($\ell=1$) & \iobs ($\ell=2$)\\ 
      \noalign{\smallskip}
      \hline 
      \noalign{\smallskip}
        20 & 16 & 13 &  494.564 & 285.536\\
        25 & 20 & 16 &  614.458 & 354.757\\
        26 & 21 & 17 &  644.432 & 372.063\\
        31 & 25 & 20 &  764.326 & 441.284\\
        32 & 26 & 21 &  794.300 & 458.589\\
        37 & 30 & 24 &  914.194 & 527.810\\
        38 & 31 & 25 &  944.168 & 545.115\\
        40 & 32 & 26 &  974.141 & 562.421\\
        42 & 34 & 27 & 1034.089 & 597.031\\
        43 & 35 & 28 & 1064.062 & 614.336\\
        44 & 36 & 29 & 1094.036 & 631.642\\
        45 & 36 & 29 & 1094.036 & 631.642\\
        46 & 37 & 30 & 1124.009 & 648.947\\
        47 & 38 & 31 & 1153.983 & 666.252\\
        48 & 39 & 31 & 1183.957 & 683.557\\
        49 & 40 & 32 & 1213.930 & 700.863\\
      \noalign{\smallskip}
      \hline 
      \end{tabular}
      \label{tab:ident_mixed}
   \end{center}
\end{table}
%
\subsection{Scenario 2 \label{ssec:scen2}}

The second scenario implies different radial and azimuthal orders. 
To check it, the $\ell=1$ modes with ($n, m$)=(24,+1), (22,0) and
(19,-1) have been arbitrarily selected. Their corresponding frequencies are
\eqn{\sigma_{19,-1}=16.727,\;\;\sigma_{22,0}=13.493,\;\;\sigma_{24,+1}=10.979,
     \nonumber\label{eq:game_mixed}}
for which the corresponding frequency ratios are
\eqn{\frac{\sigma_{+1}}{\sigma_{0}}=0.8137,\;\;\frac{\sigma_{0}}{\sigma_{-1}}=0.8066,\;\;
      \frac{\sigma_{+1}}{\sigma_{-1}}=0.6563\, .
      \nonumber\label{eq:game_mixed_ratios}}
In this case, the application of the FRM to these frequency ratios yields the
list of possible ($n$, ${\cal I}$) identifications given in 
Table~\ref{tab:ident_mixed}. Once again, the solution is
not predicted by the FRM. 
Moreover, the \iobsrot\
predictions for $\ell=1$ modes are out of the error limit. However,
as discussed in Scenario~1, when considering $\ell=2$ modes, the
theoretical \vaiss\ integral is predicted, at least by one of the
sets: $(n_1, n_2, n_3)=(47, 38, 31)$.
\begin{figure*}
 \begin{center}
   \includegraphics[width=9cm]{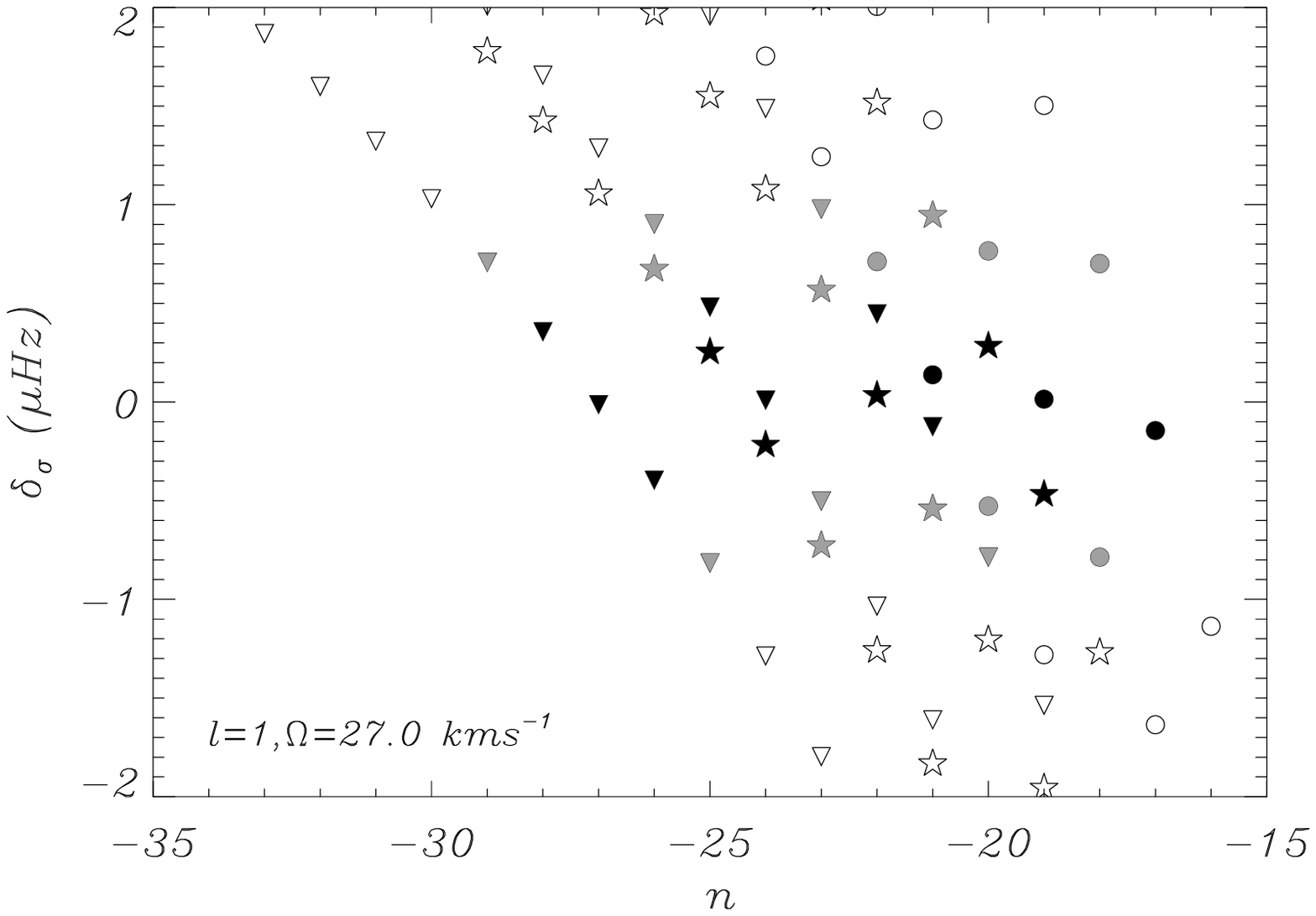}\hspace{-0.5cm}
   \includegraphics[width=9cm]{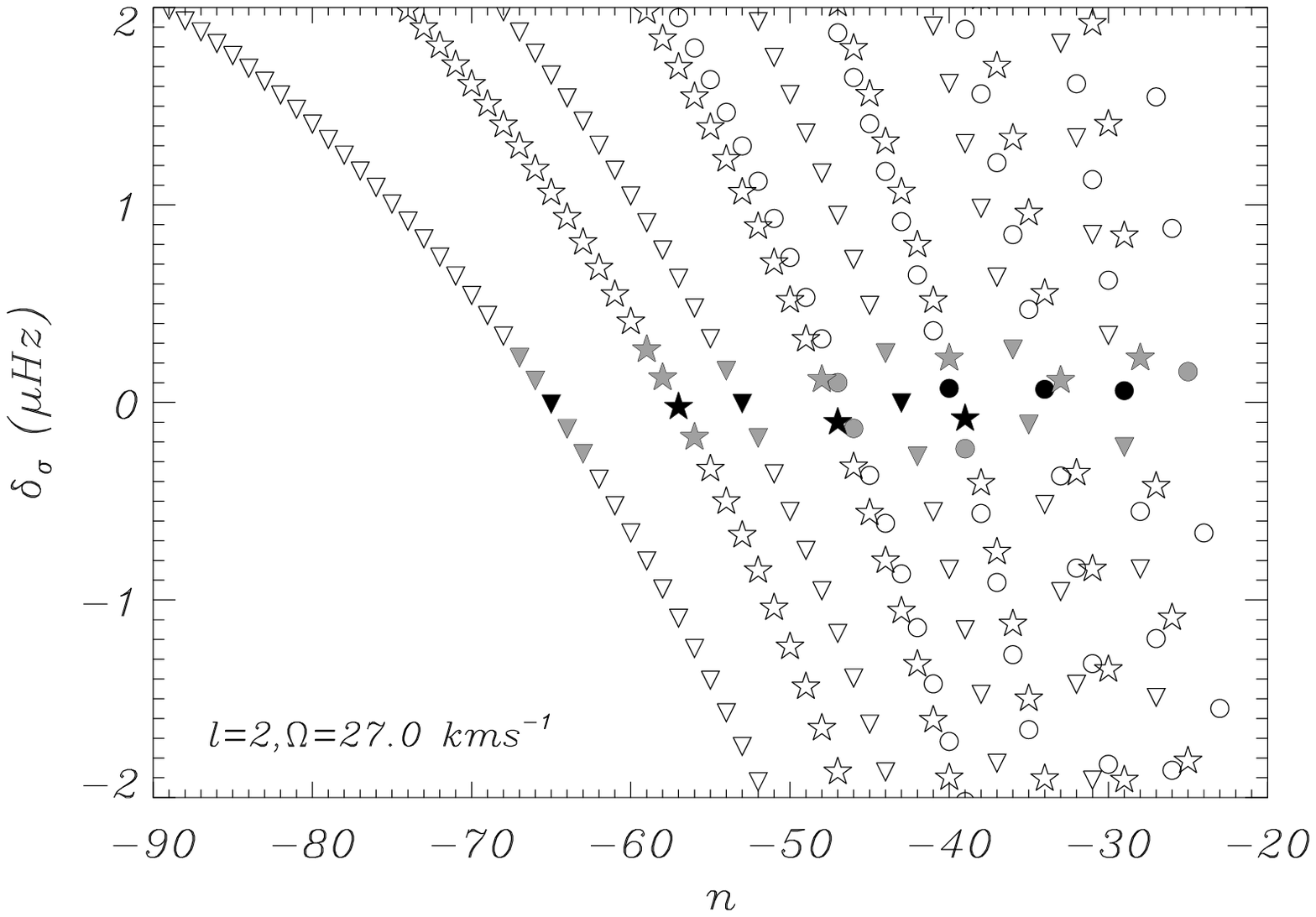}\hspace{-0.5cm}   
   \includegraphics[width=9cm]{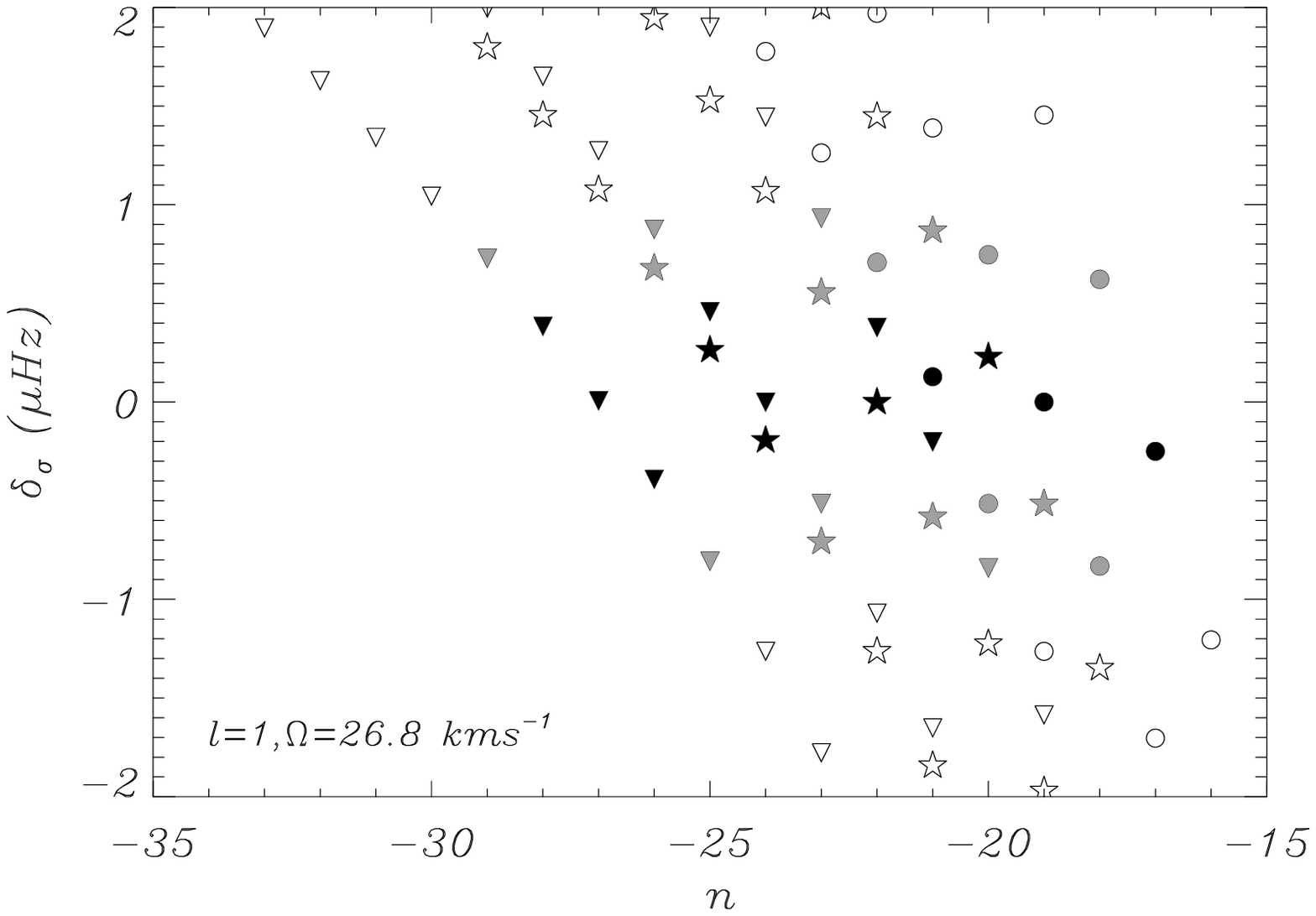}\hspace{-0.5cm}
   \includegraphics[width=9cm]{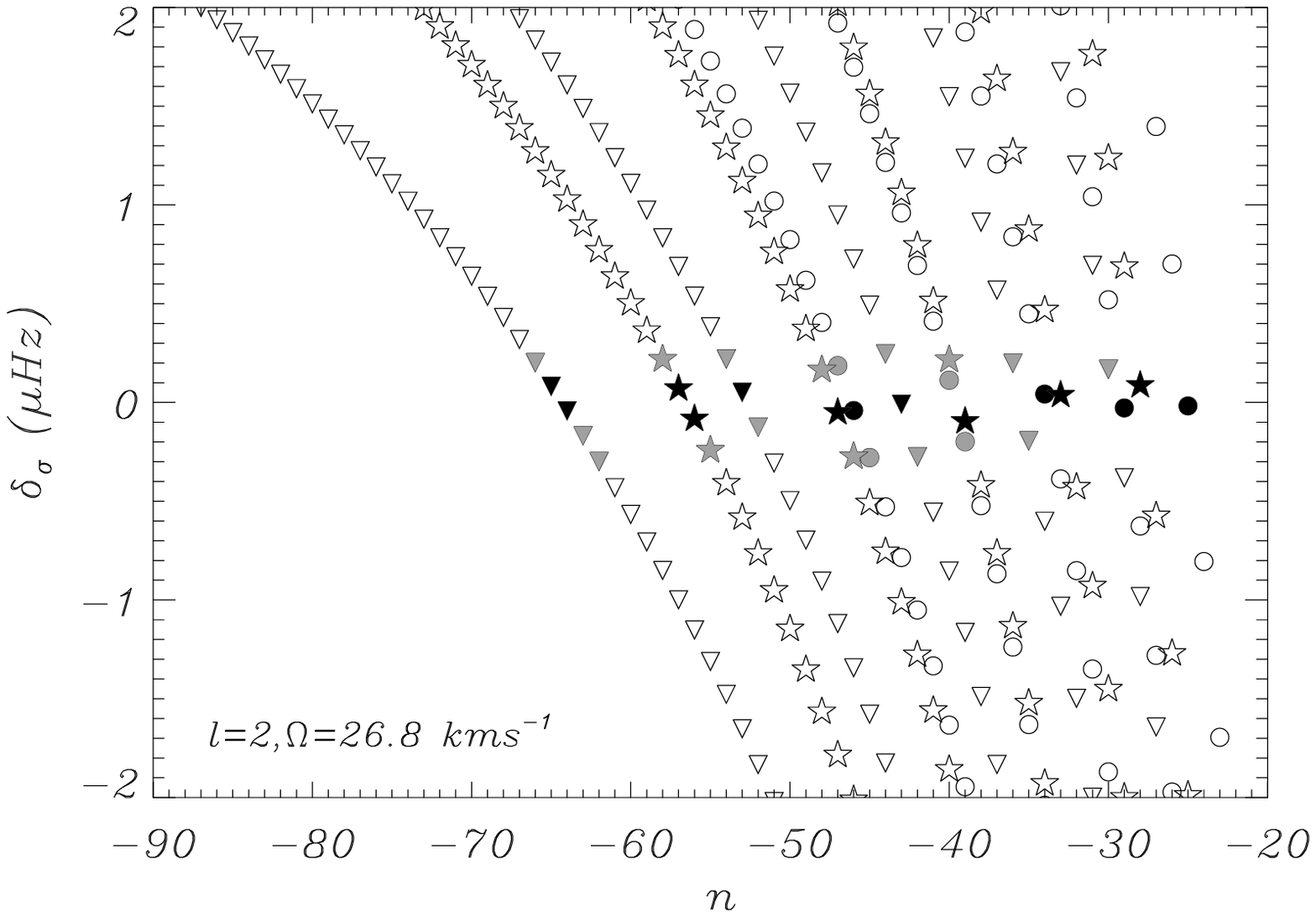}\hspace{-0.5cm}   
   \includegraphics[width=9cm]{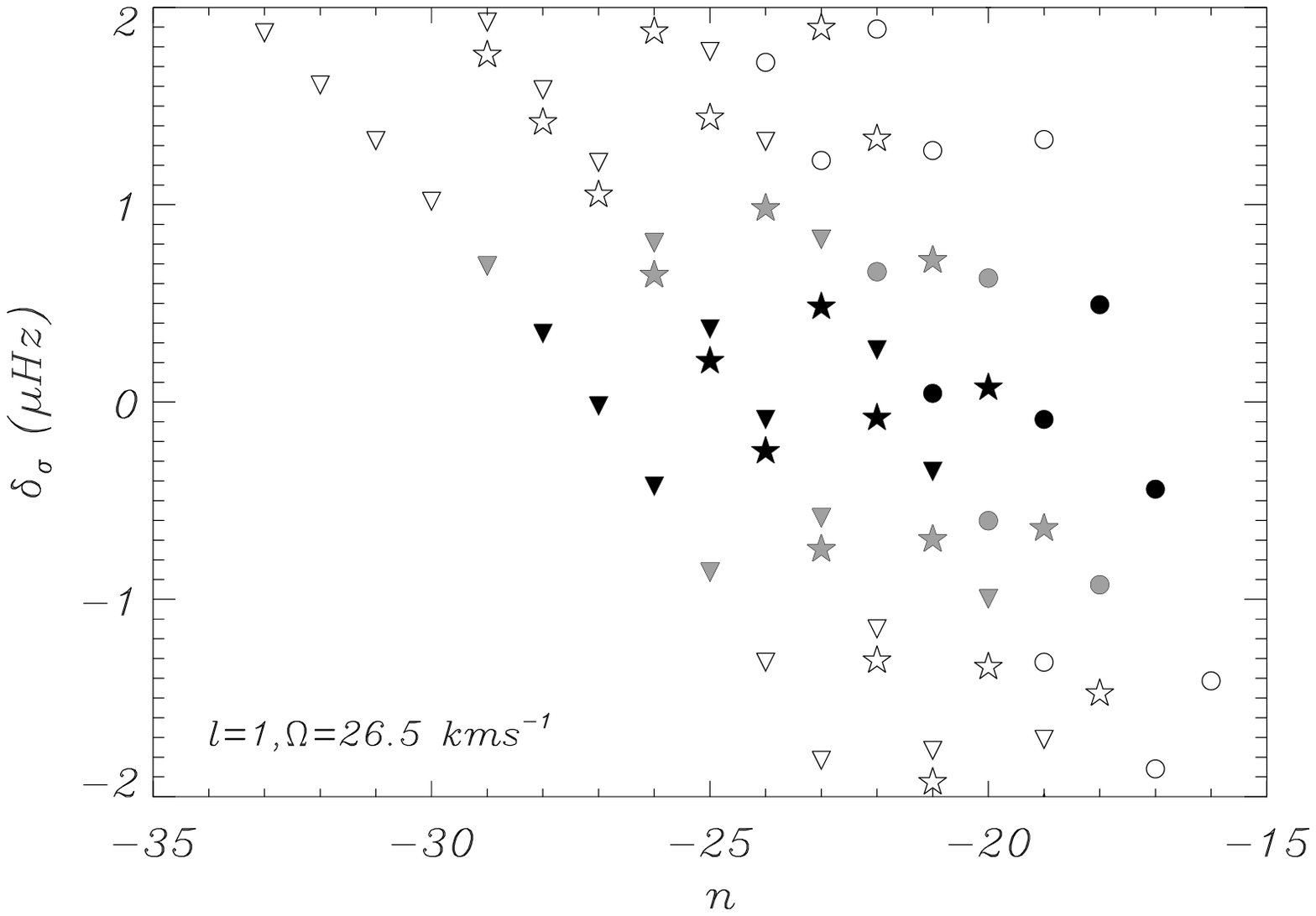}\hspace{-0.5cm}
   \includegraphics[width=9cm]{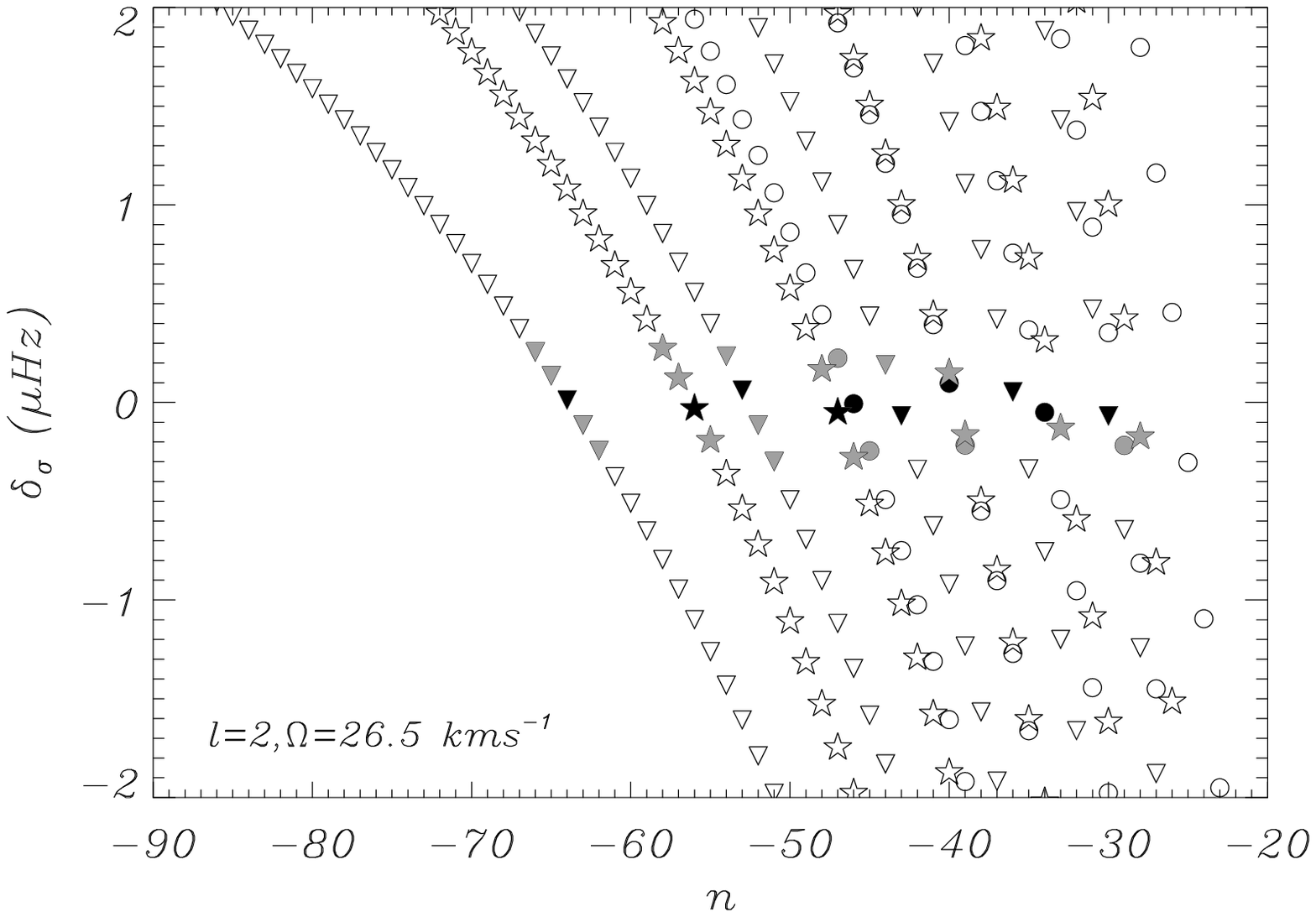}   
   \caption{Frequency differences between the components of the 
            \emph{observed} modes proposed in Scenario 3 (Sec.~\ref{ssec:scen3}) and 
	    those of theoretical multiplets as a function of the radial order $n$. 
	    Left and right panels show frequency
	    differences calculated considering $\ell=1$ and $\ell=2$ modes 
	    respectively. Symbols represent the frequency differences calculated
	    for the three \emph{observed} $\ell=1, m=0$ modes: circles
	    for $\sigma_{19}$, triangles for $\sigma_{22}$ and finally, stars
	    for $\sigma_{24}$. Azimuthal orders should be read as going from negative to
	    positive ($m=\pm | \ell |$) from left to right respectively. 
	    Filled circles represent modes for which $\delta_\sigma\lesssim0.5$
	    (grey) and $\delta_\sigma\lesssim0.1$ (black). }
   \label{fig:diffreq}
 \end{center} \vspace{4cm}
\end{figure*}

\subsection{Scenario 3 \label{ssec:scen3}}

The third and last scenario is constituted by three centroid 
modes ($m=0$), with different radial order. Proceeding as in 
Scenarios~1 and 2, the 
frequencies of the $\ell=1$ modes with $n=19$, 22 and 24 are
\eqn{\sigma_{19}=15,4633,\;\;\sigma_{22}=13.4928,\;\;\sigma_{24}=12.4232,
     \nonumber\label{eq:game_centroid}}
and their frequency ratios are
\eqn{\frac{\sigma_{24}}{\sigma_{22}}=0.9207,\;\;\frac{\sigma_{22}}{\sigma_{19}}=0.8726,\;\;
      \frac{\sigma_{24}}{\sigma_{19}}=0.8034\, .
     \nonumber \label{eq:game_centroid_ratios}}
The ($n$, ${\cal I}$) possible identifications given by the FRM applied to these
frequencies are listed in Table~\ref{tab:ident_centroid}. In contrast to the previous
scenarios, when considering \emph{centroid} modes 
the solution is found by the FRM. Not only the \emph{observed}
radial order set is included in the list of possible solutions (and
thereby the mode degree $\ell$), but also its related \iobsrot\ value predicts
the theoretical \vaiss\ integral within the error limits.  

\subsection{Considering different rotational velocities \label{ssec:scen4}}

The previous discussion considers only one rotational velocity.
However, usually, it cannot be determined from observations neither
the specific surface rotational velocity nor the angle of inclination of the star
but the value of $\vsini$. Since frequency patterns are $\Omega$ dependent, it is plausible 
to imagine a given rotational velocity for which Eq.~\ref{eq:condtrip} holds.
To investigate this possibility, 7 models with different rotational velocities 
have been considered: three models with $\Omega<\Omega_0$, three models 
with $\Omega>\Omega_0$ and finally the model used for the theoretical test 
which rotates at $\Omega_0=26.8\,\kms$. The range of rotational velocities
has been chosen as to produce models representative of a typical \gd\ star. The 
HR diagram error boxes have been obtained from typical multicolour photometry
errors of \gds. 
\begin{table}
   \caption{Same as Table~\ref{tab:ident_triplet} for the case of 
            modes with different radial order but the
	    same ($m=0$) azimuthal order (more details in the text).}
   \begin{center}
      \begin{tabular}{rrrrr} \hline\hline
      \noalign{\medskip}
      $n_1$ & $n_2$ & $n_3$ &\iobsrot ($\ell=1$) & \iobsrot ($\ell=2$)\\ 
      \noalign{\smallskip}
      \hline 
      \noalign{\smallskip}
        24 & 22 & 19 &  674.405  & 389.368\\ 
        25 & 23 & 20 &  704.379  & 406.673\\ 
        26 & 24 & 21 &  734.353  & 423.978\\ 
        34 & 31 & 27 &  944.168  & 545.115\\ 
        35 & 32 & 28 &  974.141  & 562.421\\ 
        36 & 33 & 29 & 1004.115  & 579.726\\ 
        37 & 34 & 30 & 1034.089  & 597.031\\ 
        39 & 36 & 31 & 1094.036  & 631.642\\ 
        40 & 37 & 32 & 1124.009  & 648.947\\   
      \noalign{\smallskip}
      \hline 
      \end{tabular}
      \label{tab:ident_centroid}
   \end{center}
\end{table}

We are looking for any trend indicating that the \emph{observed} frequencies
$\sigma_{19}$, $\sigma_{22}$, and $\sigma_{24}$ may be misidentified with a rotational
triplet or with components of any $\ell=1, 2$ multiplets. To do so, we have
calculated the frequency difference $\delta_\sigma$ between the \emph{observed} modes and those
of theoretical multiplets for the 7 models considered here. In Fig.~\ref{fig:diffreq},
such differences are shown for the reference model (central panels), the fastest one
(top panels) and the slowest one (bottom panel). In the case of
$\ell=1$ triplets (left panels), the smallest $|\delta_\sigma|$ values 
vary from 0.1 (black filled circles) to 0.5 (grey filled circles) approximately, 
for a set of 15-20 modes. 
As expected, for $\ell=2$ modes (right panels)
$\delta_\sigma$ decreases, and the number of modes with frequencies close
to the observed ones increases. In this case, the smallest $|\delta_\sigma|$ values  
vary from $0.001\,\muHz$ (black filled circles) to $0.3\,\muHz$ (grey filled circles) 
approximately, 
for a set of 35-40 modes. Discarding the exact solution for $\sigma_{19}$, $\sigma_{22}$, 
and $\sigma_{24}$ for which $\delta_\sigma=0$, these frequency differences exceed in 
at least three orders of magnitude ($\ell=2$ results)\footnote{For $\ell=1$, $\delta_\sigma$
differences reach five orders of magnitude that of the observational frequency error.}
the observational frequency error typical for \gds. 

\begin{figure}
 \begin{center}
   \includegraphics[width=9cm]{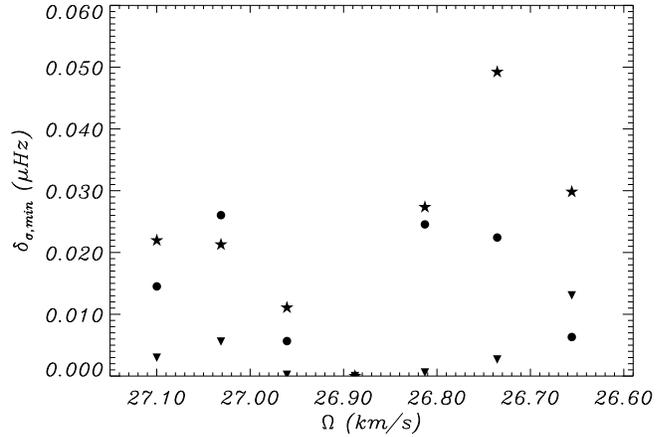}
   \caption{Variation of the minimum frequency differences considered in Fig.~\ref{fig:diffreq}
            with the rotational velocity. Same symbol representation as in Fig.~\ref{fig:diffreq} is
	    adopted. }
   \label{fig:difmin}
 \end{center} 
\end{figure}

In order to consider the smallest differences
for the 7 seven models, the minimum $\delta_\sigma$ value found for each model is depicted
in Fig.~\ref{fig:difmin} as a function of the rotational velocity. 
Although the mode degree is not here specifically identified, it is found that
most of selected modes are $\ell=2$, result somehow expected since, according to 
Fig.~\ref{fig:diffreq}, the smallest differences corresponds to $\ell=2$. 
Obviously, the exact solution 
($\delta_{\sigma,{\rm min}}=0$ at $\Omega=\Omega_0$) correspond to $\ell=1$ modes.
As can be seen, the $\delta_{\sigma,{\rm min}}$ distribution
does not show any specific trend with the rotational velocity and, quantitatively, the
results obtained are similar to those shown in Fig.~\ref{fig:diffreq}. 
This result does not exclude the possibility of misidentification, however it
shows it is unlikely for the range of models representative of the
observed star.

This theoretical test shows that, provided the
mode degree $\ell$ is known, the FRM may be discriminating with respect to the
multiplet-like structures. In such cases, coherent FRM solutions are restricted to
centroid mode ($m=0$) frequencies\footnote{Such result is obviously restricted to stars
fulfilling the conditions and errors discussed in Sect.~\ref{sec:errors}.}.
When no additional information on $\ell$ is given, the FRM does not
discriminate multiplet-like structures. Nevertheless, as shown in this 
exercise, such a situation is unlikely. 
\section{A particular case: the \gd\ star HD\,48501\label{sec:applic}}

The \gd\ star \object{HD\,48501} is particularly suitable to illustrate the previous
theoretical exercise applied to a real star. Having a $\vsini\sim30\,\kms$, it can be considered 
as a slow rotator, fulfilling thus one of the conditions discussed in Sect.~\ref{sec:errors}. 
According to a recent frequency analysis 
reported by \citet{AertsCuypers04} the star pulsates in at least three modes. 
In this 
paper, a discussion has been opened on whether these frequencies could belong 
to a $\ell=1$ triplet-like
structure or to the period spacing expected for high-order gravity modes.

\subsection{Fundamental parameters\label{ssec:fundparam}}

The stellar fundamental parameters of \hd\ have been determined by
applying \emph{TempLogG} \citep{templogg} to the Str\"omgren--Crawford
photometry listed in the Hauck-Mermilliod catalogue
\citep{Hauck98}. In this catalogue, no $\beta$ index value is given
for \hd, which was instead obtained from \citet{Handler99}. The code
classifies this object as a main sequence star in the spectral range
region F0-G2.  The resulting physical parameters are listed in
Table~\ref{tab:params}.
\begin{table}
     \caption{Fundamental parameters of \hd\ taken
     from the literature. From left to right:
     effective temperature (in K); surface
     gravity (cgs, in a logarithmic scale); 
     metallicity and finally the projected rotational
     velocity (in $\kms$). Superindex indicates data
     were obtained from the following references:
     ($a$)~\citet{templogg}; ($b$)~\citet{Dupret02};
     ($c$)~\citet{CayreldS97}; ($d$)~\citet{AertsCuypers04}
     and finally ($e$)~\citet{Royer02}.}
     \begin{center}
     \vspace{1em}
     \renewcommand{\arraystretch}{1.2}
     \begin{tabular}{cccc}
     \hline\hline
         $\teff$ & $\logg$  & [Fe/H]    & $\vsini$  \\
           (K)   &   (dex)  & (dex)     & $(\kms$)   \\
     \hline
    {\bf 6984}$^a$ & {\bf 3.92}$^a$ & {\bf-0.12}$^a$ &  29$^d$    \\
        $7079^b$ & $4.49^b$ & $-0.10^b$ &  47$^e$    \\
        $7099^c$ & $3.96^c$ & $+0.01^c$ &          \\
     \hline
     \end{tabular}
     \label{tab:params}
     \end{center}    
\end{table}

\citet{Dupret02} gives also physical parameters for \hd\ based on
the Geneva photometry of \citet{AertsCuypers04} and the calibrations given by
\citet{Kunzli97} for the Geneva photometry of B to G stars. In his
study, theoretical stellar models for these stars do not take into account
their surface gravity determinations, for being too high, 
corresponding to models below 
the ZAMS. Physical parameters
were also retrieved from the fifth edition of the catalogue
of [Fe/H] determinations by \citet{CayreldS97}, which includes [Fe/H] and
atmospheric parameters ($\teff$, $\logg$) obtained from high
resolution spectroscopic observations and detailed analysis, most of
them carried out with the help of model-atmospheres.  The
aforementioned physical parameters are listed in Table~\ref{tab:params}.

Values for the $\vsini$ of were found in \citet{Royer02}, computed from spectra 
collected at Observatoire de
Haute-Provence (OHP) and by \citet{AertsCuypers04}, derived from a
cross-correlation function analysis of spectroscopic measurements with
the CORALIE spectrograph. Very recently, \citet{Mathias04} have
published the results of a two-year long high-resolution spectroscopy
campaign, monitoring 59 \gd\ candidates. In this campaign, more than
60\% of the stars presented line profile variations which can be
interpreted as due to pulsation. 
\begin{table}
  \begin{center}
    \caption{Photometric data for the \gd\ star \hd. Observed frequencies, 
             in both cycles per day and $\muHz$, are taken from
	     \citet{AertsCuypers04}.}
    \vspace{1em}
    \renewcommand{\arraystretch}{1.2}
    \begin{tabular}[h]{cccc}
      \hline\hline
                                     & ($c/d$) & ($\muHz$) \\
      \hline       
                       $f_{\rm I}$   & 1.095   & 12.663    \\
	               $f_{\rm II}$  & 1.199   & 13.880    \\		       
                       $f_{\rm III}$  & 1.290  & 14.937    \\

      \hline
      \end{tabular}
    \label{tab:obsfreq}
  \end{center}
\end{table}
%

\subsection{Locating HD\,48501 in the HR diagram\label{ssec:HR_hd48501}}

\begin{figure*}
 \begin{center}
   \includegraphics[width=13.5cm]{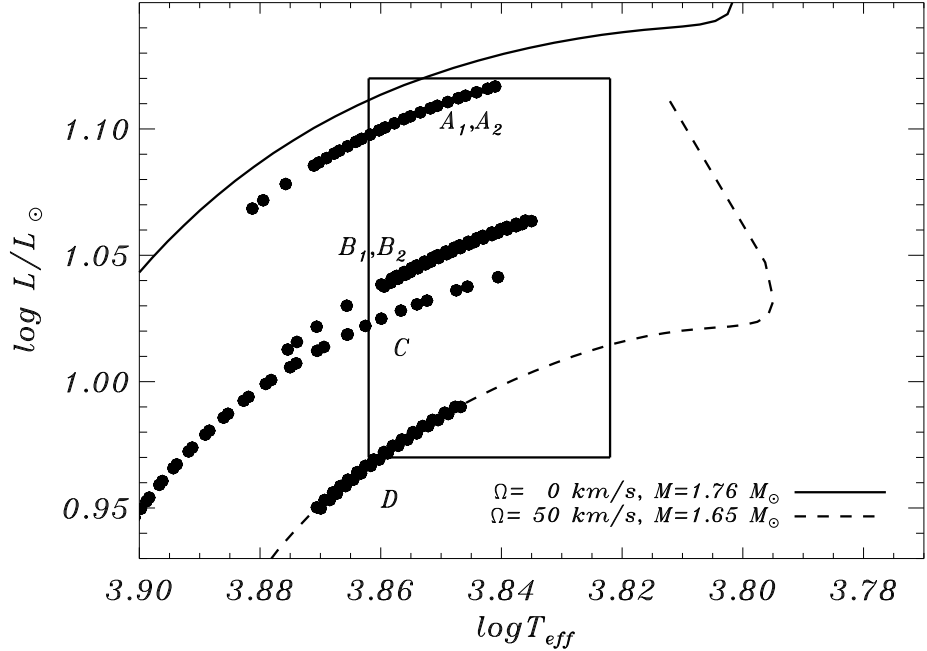}
   \caption{HR diagram containing evolutionary tracks and models loci
            considered in this work. The continuous line represents the
	    evolutionary track of a $1.76\,\msol$ non rotating star model. 
	    Respectively, the
	    dashed line represents an evolutionary track of an $1.65\,\msol$
	    pseudo-rotating star model, considering a rotational velocity of 
	    $50\,\kms$. $A$, $B$ and $D$ correspond to the
	    location of $1.75,\msol$, $1.70,\msol$ and $1.65\,\msol$
	    pseudo-rotating models respectively (a solar metallicity is 
	    considered). Subindexes 1 and 2 represent
	    values of rotation velocities at the stellar surface of $30\,\kms$
	    and $50\,\kms$ respectively. $C$ represents models of $1.50\,\msol$,
	    with a metallicity of $Z=0.07$ and a rotational velocity of 
	    $50\,\kms$. Models are displayed with filled circles. Finally, the 
	    error box considered for \hd\ is also depicted.}
   \label{fig:2gdors_diagHR}
 \end{center}
\end{figure*}

As usual, the modelling of any star is conditioned by its stellar parameters,
delimiting its position in the HR diagram with the corresponding
photometric errors. In the present case, the photometric error box 
(see Fig.~\ref{fig:2gdors_diagHR}) has been delimited from 
fundamental parameters (Table~\ref{tab:params}) obtained from the 
Str\"omgren--Crawford photometry (boldface). For these values, 
a standard accuracy of $0.2\,\mbox{dex}$ for the surface 
gravity and $200\,\mbox{K}$ in 
$\teff$ has been considered. In addition, a typical correction for the effect of 
rotation is included as well. For main sequence intermediate mass
stars, this represents $\pm200\mbox{K}$ approximately 
\citep[details in][]{MiHer99, Pe99}, 
and more recently in \citet{Sua02aa}.
\begin{figure*}
 \begin{center}
   \includegraphics[width=13.5cm]{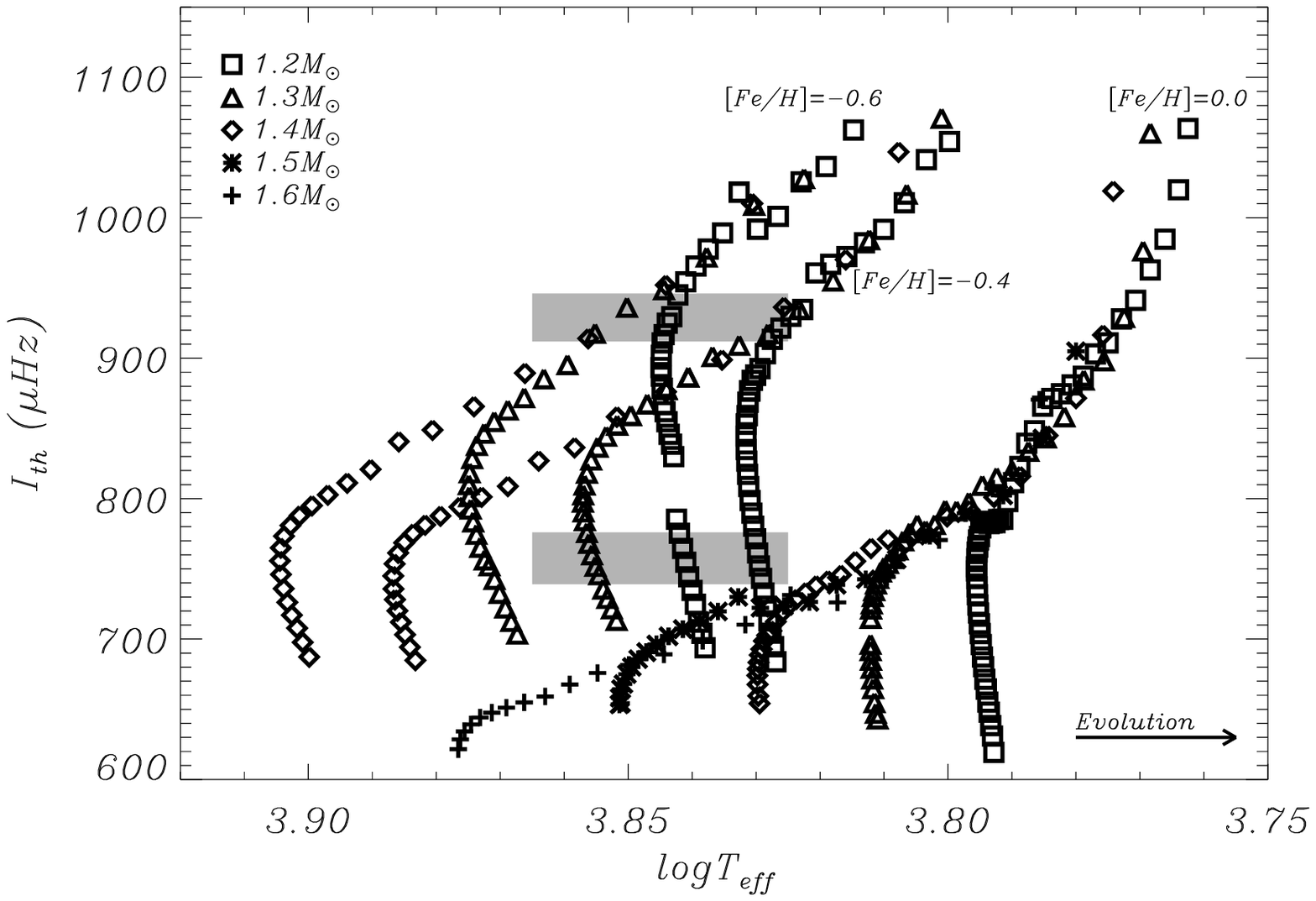}
   \caption{Theoretical \vaiss\ frequency integral as a function of the effective 
            temperature for representative models of \hd\ with masses in the range of 
	    $M=1.2$--1.4$\msol$, for $[Fe/H]=-0.4$ and -0.6, and $M=1.2$--1.6$\msol$
	    for $[Fe/H]=0.0$, -0.4 and -0.6. Shaded areas represent uncertainty
	    (${\cal I}_{\mathrm{obs}}$,$\teff$) boxes of sets $t_1$--$t_3$
	    and $t_4$, $t_6$ as given in Table~\ref{tab:hdchains}.}
   \label{fig:intI-teff}
 \end{center}
\end{figure*}
In the same figure, several evolutionary tracks are drawn to illustrate
the wide range of different models that can be considered as representative
of \hd. In particular, the location in the HR diagram of evolutionary tracks can be 
modified significantly using different rotational velocities. This becomes critical 
when different values for the rotational velocity and the metallicity are 
taken into account.
\begin{figure*}
 \begin{center}
   \includegraphics[width=8.5cm]{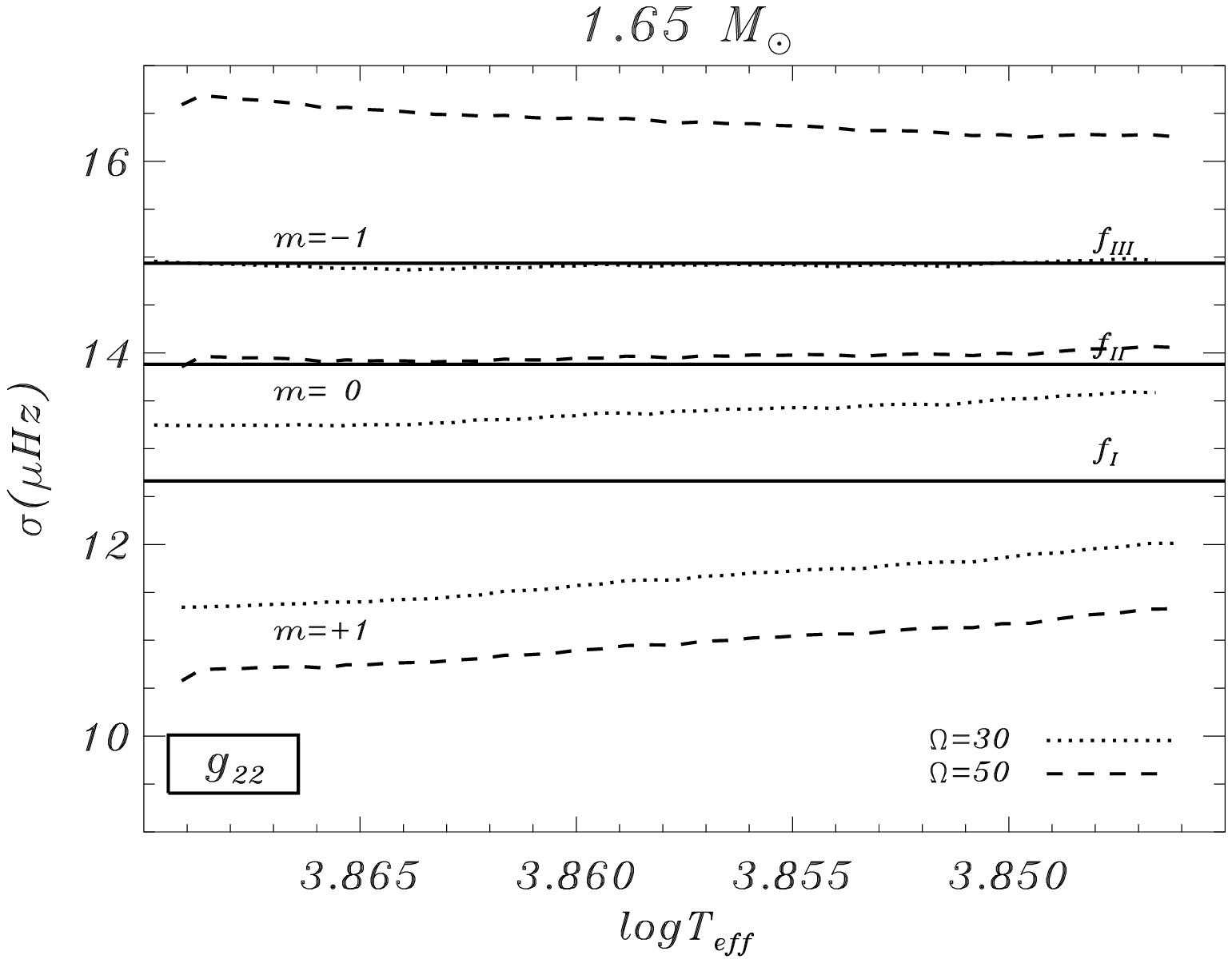}
   \includegraphics[width=8.5cm]{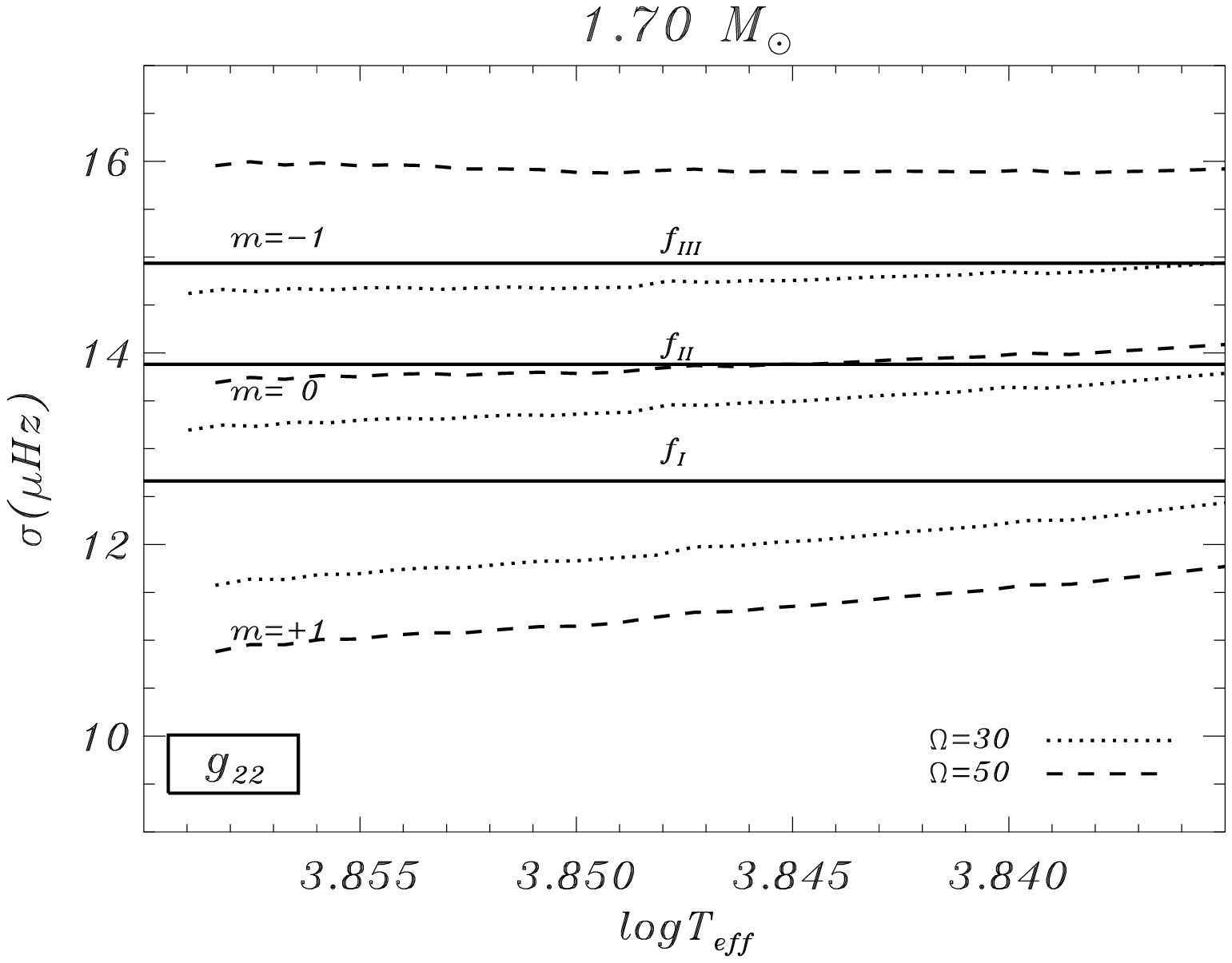}
   \caption{Evolution of adiabatic oscillation frequencies for the
            $1.65\,\msol$ and $1.70\,\msol$ models within the 
	    \hd\ photometric error box (left and right panels respectively).
	    Only the three components of the $g_{22}$ triplet are displayed. 
	    Dotted lines represent
	    the frequencies obtained from models with a rotational velocity of 
	    $30\,\kms$, and dashed lines correspond to those obtained from models
	    with a rotational velocity of $50\,\kms$. Finally, the observed 
	    frequencies are represented by continuous horizontal lines.}
   \label{fig:evolfreq}
 \end{center}
\end{figure*}
At present,
there is no additional information constraining the angle of inclination
of \hd. Therefore, a wide range of rotational velocities should
be considered. However, in the present case, in order to respect 
the hypothesis of applicability of the FRM, we restrict our modelling
to low rotational velocities. In Fig.~\ref{fig:2gdors_diagHR},
the upper continuous line and the lower dashed line represent, respectively, 
the evolutionary tracks of a $1.76\,\msol$ non rotating star model and a 
$1.65\,\msol$ pseudo-rotating star model. These tracks illustrate somehow 
the limits for the mass and rotational velocities predicted by the modelling.

Four different masses have been considered: from bottom to top, 
$1.65\msol$ ($D$ models), $1.70\,\msol$ ($B$ models) and $1.75\msol$ 
($A$ models) computed with solar metallicity. To cover the observed
$\vsini$ found in the literature (see Table~\ref{tab:params}), the
following rotational velocities have been used: $30\,\kms$ ($A_1$, $B_1$) 
and $50\,\kms$ ($A_2$, $B_2$), corresponding to an angle of inclination 
of $90^\circ$ and $37^\circ$ respectively.  

\subsection{Applying the FRM to the observed frequencies\label{ssec:applicFRM-hd48501}}

The three observed frequencies (Table~\ref{tab:obsfreq}) of \hd\ have been taken from 
\citet{AertsCuypers04}. The corresponding frequency
ratios are
\eqn{\frac{f_{\rm I}}{f_{\rm II}}=0.9123,\;\;\frac{f_{\rm II}}{f_{\rm III}}=0.9292,\;\;
      \frac{f_{\rm I}}{f_{\rm III}}=0.8478\, .
     \nonumber \label{eq:hd_ratios}}
All possible natural number ratios up to $n=60$ are then
calculated. As described in Sect.~\ref{sec:errors}, we assume
an error of $\pm1.3\cdot10^{-2}$ for the calculation of 
possible $n$ sets, instead of the standard error of 
$\pm5\cdot10^{-3}$ used under the assumption of 
non-rotation (cf. \paperI). The results are detailed
in Table~\ref{tab:hdchains}, for which only dipoles
($\ell =1$) and quadrupoles ($\ell =2$) have been 
considered up to $n\lesssim50$.

Within the aforementioned errors, seven possible $n$ sets are 
obtained. Although, most of them (5 of 7)
are predicted as $\ell=2$ modes, the $t_4$ and $t_6$ are
predicted as $\ell=1$ modes. This latter possibility would
thus support the results obtained with multicolour photometry
given by \citet{AertsCuypers04}. 
The corresponding \iobsrot\ integrals 
(see Table~\ref{tab:hdchains}) are compared with theoretical 
predictions in the \iteo\---$\teff$ diagram given in Fig.~\ref{fig:intI-teff}.
When analysing such comparison for $\ell=1$ possibilities (shaded
boxes), no successful results are afforded, i.e. the models 
fulfilling the $(n_i,{\cal I}$) predictions are not consistent
with the observational constraints. In particular, the observed metallicity 
and the corresponding \iteorot\ predictions are in clear 
dissonance. In particular, the FRM would predict
models with a metallicity
in the range of [Fe/H]=[-0.4,-0.6]. 
In contrast, when considering possible sets with $\ell=2$, 
the FRM predicts (according to the \iobsrot\ integrals) models 
with the observed metallicity [Fe/H]$\geq0.1$, in particular
for $t_1$, $t_2$ and $t_3$ sets. Such solutions would correspond
to models with masses from $1.4$ to $1.6\,\msol$ in the main
sequence.

\subsection{The observed frequencies as belonging to a rotational split $\ell=1$ triplet
           \label{ssec:triplet-hd48501}}
	   
The previous results suggest, if $\ell=1$ confirmed, the alternative of considering the
observed frequencies as belonging to one or more multiplet-like
structure split by rotation.

In Fig.~\ref{fig:evolfreq}, the theoretical evolution of adiabatic
frequencies is given for two representative models of \hd. The region 
where the observed frequencies (continuous horizontal lines) can be identified 
is around the triplet $g_{22}$, that is a $n=22$, $g$ mode. This figure shows 
that theoretical predictions fit better for higher masses (right panel) and for
evolved models (lower effective temperatures). 
\begin{table}
  \begin{center}
  \caption{List of selected natural numbers sets associated to the 
           observed frequency ratios of \hd. For each set, the 
	   corresponding \emph{observed} \iobsrot\ is given in $\muHz$.
	   Finally, an estimate of the spherical order $\ell$ is also
	   given in column 4. }
  \begin{tabular}{cccccc} \hline
  \noalign{\medskip}
& $n_1$ & $n_2$ & $n_3$ & $\ell$ & ${\cal I}_{\mathrm{obs}}$\\ 
   \noalign{\smallskip}
\hline 
   \noalign{\smallskip}
$t_{1}$& 38 &  41 &  45 & 2 & 739\\
$t_{2}$& 39 &  42 &  46 & 2 & 757\\
$t_{3}$& 40 &  43 &  47 & 2 & 776\\
$t_{4}$& 27 &  29 &  32 & 1 & 912\\
$t_{5}$& 27 &  29 &  32 & 2 & 527\\
$t_{6}$& 28 &  30 &  33 & 1 & 946\\
$t_{7}$& 28 &  30 &  33 & 2 & 546\\
\hline
  \end{tabular}
  \label{tab:hdchains}
  \end{center}
\end{table}
%
\begin{figure*}
 \begin{center}
   \includegraphics[width=9cm]{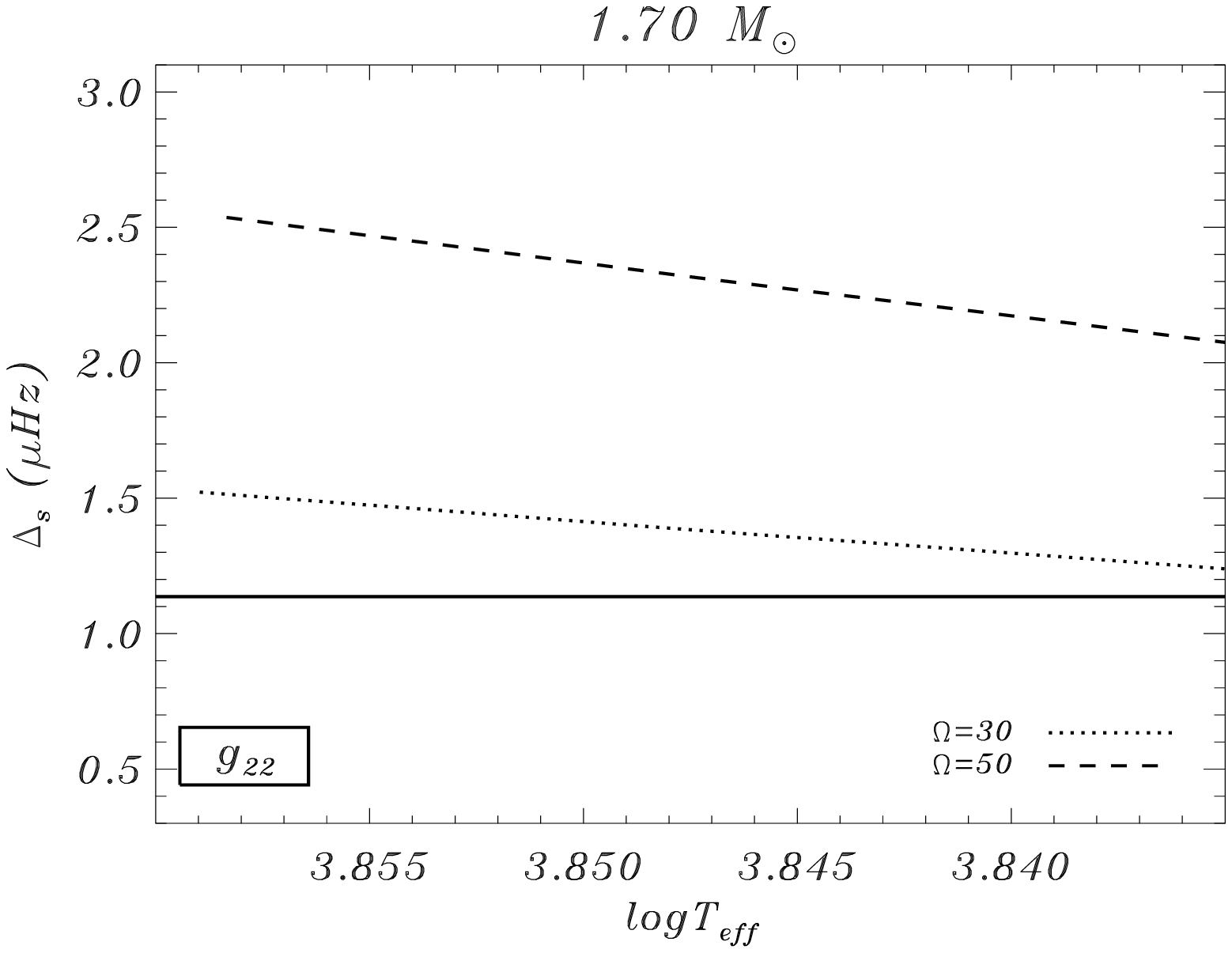}\hspace{-0.5cm}
   \includegraphics[width=9cm]{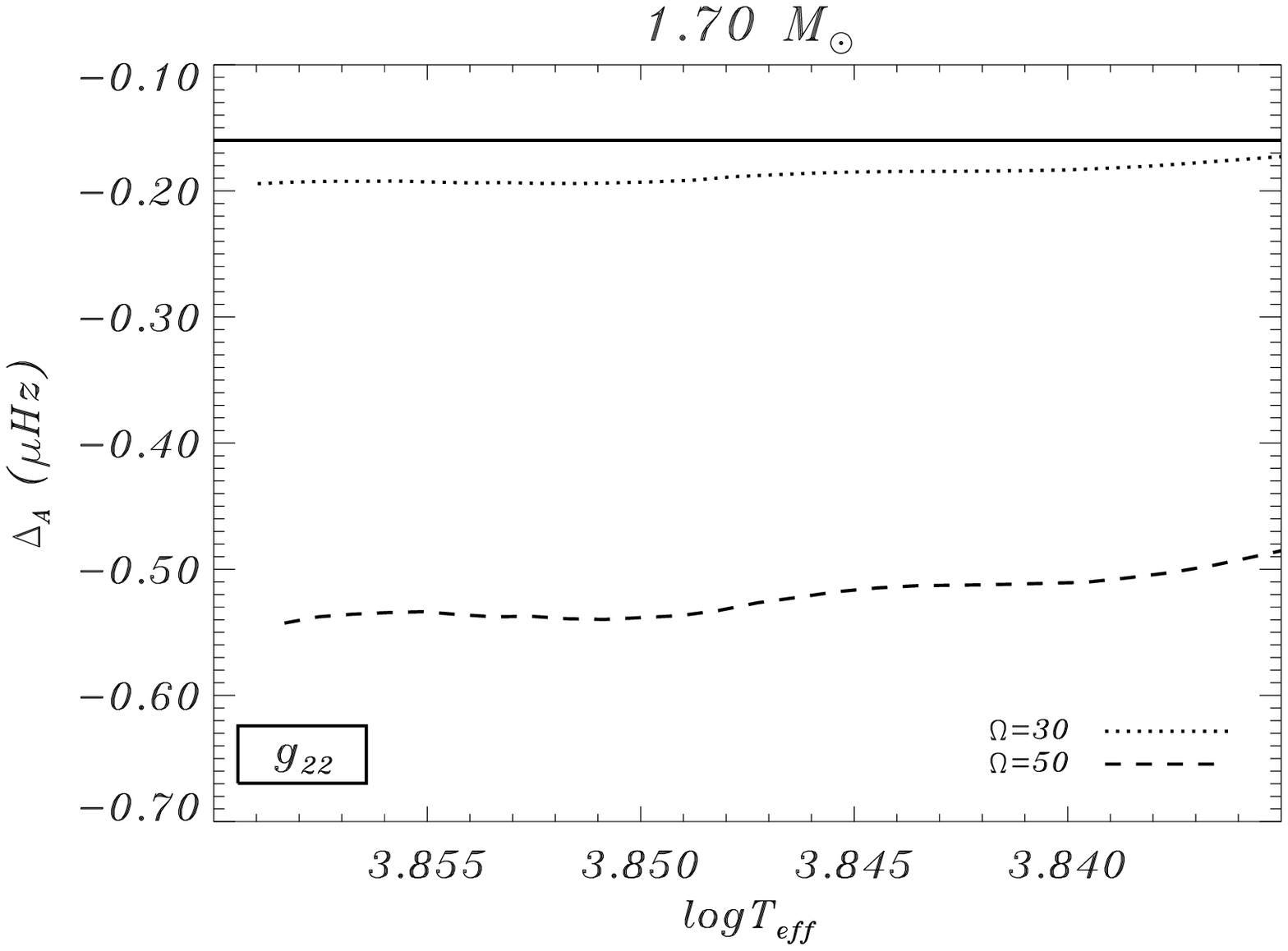}
   \caption{Mean splitting (left panel) and asymmetry (right panel) of the 
            selected $(\ell,n)=(1,22)$ triplet of the $1.70\,\msol$ models 
	    as calculated in 
	    Eqs.~\ref{eq:defsplitting} and \ref{eq:defasym} 
	    respectively. Solid lines represent observed
	    values. Dotted and dashed lines represent theoretical
	    values obtained from models with a rotational velocity
	    of $30$ and $50\,\kms$ respectively.}
   \label{fig:DeltaSA}
 \end{center}
\end{figure*}
Two rotational
velocities are considered: $30\,\kms$, and $50\,\kms$, represented by
dotted and dashed lines respectively. For the \emph{low-mass} model
(Fig.~\ref{fig:evolfreq}, left panel), the predicted triplet structure is 
clearly shifted with respect to observed frequencies for both rotational
velocities. In the case of the $1.70\,\msol$ model, the predicted 
$m=0$ frequency obtained when considering $\Omega=50\,\kms$ is 
\emph{centred} with respect to the observed ones. However the
$m=\pm1$ branches are still far from the observations. In contrast,
when considering the $\Omega=30\,\kms$ rotational velocity, 
theoretical frequencies tend to the observed ones for the low
effective temperatures. Such results would thus suggest that 
$1.70\,\msol$ pseudo-rotating models, with effective
temperatures near the lower limit of the error box, and with a rotation
velocity around $30\,\kms$ may be considered as representative for \hd. 
In addition,
the rotational velocity confirms the spectroscopic $\vsini$ observation.  


In order to confirm the triplet possibility, it is worth
to analyse the behaviour of the average splitting predicted by models. 
For the rotational 
velocities considered here, the frequency
structure of a rotationally split $\ell=1$ can be studied
by using the mean splitting form
\eqn{\Delta_{\mathrm{s}}\equiv\frac{1}{2}\dst\left(\sigma_{+1}-\sigma_{-1}\right)\,.
     \label{eq:defsplitting}}
In Fig.~\ref{fig:DeltaSA} (left panel) the evolution of the theoretical splitting 
($\ell=1, g_{22}$) is shown for the model giving the best frequency results ($1.70\,\msol$). 
The solid line represents the observed average spacing, which value is 
$\Delta_\mathrm{s}^{\mathrm{obs}}=1.14\,\muHz$. When
considering a rotational velocity of $30\,\kms$ (dotted line), the observed mean splitting
is reproduced 
($\Delta_\mathrm{s}^{\mathrm{th}}-\Delta_\mathrm{s}^{\mathrm{obs}}\sim10^{-3}-10^{-2}$)
by models in the region of low temperatures.

In addition,
the study of the observed asymmetry may furnish additional information
to identify the corresponding triplet. 
An equivalent form to Eq~\ref{eq:defsplitting}
can be used to analyse the corresponding $\ell=1$ triplet asymmetry:
\eqn{\Delta_{\mathrm{A}}\equiv\sigma_{+1}+\sigma_{-1}-2\sigma_0\,,
     \label{eq:defasym}} 
where the subscripts $+1$, $-1$ and 0 indicate the corresponding values
of the azimuthal order $m$.
The evolution of theoretical $\Delta_\mathrm{A}$ values as obtained from models
rotating with $50\,\kms$ and $30\,\kms$ 
are compared with the observed one $\Delta_\mathrm{A}^{\mathrm{obs}}=-0.16$ in 
Fig.~\ref{fig:DeltaSA} (right panel). 
Our \emph{best} predicted asymmetry is obtained for models
with rotational velocities of $30\,\kms$. In the region of
low temperatures these models show $\Delta_{\mathrm{th}}^{\mathrm{obs}}$ approaching
to the observed $\Delta_\mathrm{A}^{\mathrm{obs}}$ value
in approximately $10^{-2}$ orders of magnitude.

The \emph{best} results for $\Delta_\mathrm{A}$ and $\Delta_\mathrm{s}$ are given
for a $1.70\,\msol$ model with a rotational velocity of 
$\Omega=28.3\,\kms$, a radius of $2.43\,\rsol$, and presenting an effective 
temperature of 6839.24 K ($\log\teff=3.835$) and $\logg=3.896$.
Its central hydrogen fraction, $X_c=0.26$, shows an evolutionary stage
corresponding to a main sequence star with 1.4 Gyr.

\section{Conclusions\label{sec:conclu}}

In the present work, the Frequency Ratio Method (FRM) proposed
by \citet{Moya05frmI} is revisited in terms of the applicability
to rotating \gds. The accuracy of the FRM in presence of 
rotation has been exhaustively examined. To do so, the following
aspects have been considered: the validity of a perturbative approach to 
compute adiabatic oscillation frequencies of $g$ modes in
asymptotic regime; the effect of rotation on the
observational \vaiss\ frequency integral and, finally, 
the problem of multiplet-like structures.
This problem concerns the possibility of disentangling whether
the observed frequencies belong to any rotational split multiplet
or to the period spacing expected for high-order gravity 
modes in asymptotic regime.

It is found that reliable results may be obtained when
objects rotate with $\Omega\lesssim70\,\kms$. In such 
cases, the error of the FRM increases on order of 
magnitude respect to the typical errors
given in \paperI.

Concerning the multiplet-like structures, 
all the possible 
\emph{confusing} scenarios have been investigated. 
Provided any additional information on the mode degree $\ell$, 
simulations indicate that the FRM would be
discriminating for $m=0$ modes. For such
cases, any misinterpretation induced by the presence
of rotationally split multiplet-like structures 
is avoided. When $\ell$ is a priori unknown, such
discrimination is not ensured. Nevertheless, we
show here that the possibility of a confusing scenario
is rather unlikely. This constitutes a
very important result, since it reinforces and extends
(up to some extent) the applicability of the FRM to 
slowly-moderately rotating
\gds.

In order to test the present results with a real star the
FRM is applied to the \gd\ star \object{HD\,48501}, which
can be considered as a slow rotator ($\vsini=29\,\kms$).
The models fulfilling the corresponding $(n_i,{\cal I}$) predictions 
were found to be inconsistent with the observational constraints
for $\ell=1$ modes, in particular with the observed metallicity. 
However, for $\ell=2$ modes, the FRM predicts models
with masses from 1.4 to $1.6\,\msol$ and radial order ranges
in the range $n=[38,47]$.

Furthermore, in contrast to predictions given by \citet{AertsCuypers04},
when analysing the observed frequencies as belonging
to a rotational induced $\ell=1$ triplet, solutions reproducing 
the observed average splitting and the triplet average asymmetry
have been found to be plausible for a $1.70\,\msol$ model.

\acknowledgements{This study would not have been possible without the financial support from 
                  the PNAYA (National Plan of Astronomy and Astrophysics, Spain), with the
		  project $AYA2003-04651$. As well, this project was also partially financed
		  by the Spanish "Consejer\'{\i}a de Innovaci\'on, Ciencia y Empresa" from the
		  "Junta de Andaluc\'{\i}a" local government, and by
		  the Spanish Plan Nacional del Espacio under project 
		  ESP2004-03855-C03-01. SMR acknowledges financial support by
		  an "Averroes" postdoctoral contract, from the Junta de Andaluc\'{\i}a local 
		  government.}


\bibliography{/home/jcsuarez/Boulot/Latex/Util/References/ref-generale}
\bibliographystyle{aa}

 \end{document}